%% file: main.tex
\documentclass[sigconf,authorversion]{acmart}
\AtBeginDocument{%
  }

\setcopyright{acmlicensed}
\copyrightyear{2026}
\acmYear{2026}
\setcopyright{cc}
\setcctype{by}
\acmConference[CHI '26]{Proceedings of the 2026 CHI Conference on Human Factors in Computing Systems}{April 13--17, 2026}{Barcelona, Spain}
\acmBooktitle{Proceedings of the 2026 CHI Conference on Human Factors in Computing Systems (CHI '26), April 13--17, 2026, Barcelona, Spain}
\acmPrice{}
\acmDOI{10.1145/3772318.3790276}
\acmISBN{979-8-4007-2278-3/2026/04}

\usepackage{graphicx}
\usepackage{multirow}
\usepackage{subcaption}
\usepackage{framed}

\begin{document}

\title[Authorship Drift: How Self-Efficacy and Trust Evolve During LLM-Assisted Writing]{Authorship Drift: How Self-Efficacy and Trust Evolve During LLM-Assisted Writing}

\author{Yeon Su Park}
\orcid{0009-0004-3071-6664}
\affiliation{
  \institution{School of Computing \\ KAIST}
  \city{Daejeon}
  \country{Republic of Korea}
}
\email{yeonsupark@kaist.ac.kr}

\author{Nadia Azzahra Putri Arvi}
\orcid{0009-0001-9527-464X}
\affiliation{
  \institution{School of Computing \\ KAIST}
  \city{Daejeon}
  \country{Republic of Korea}
}
\email{nadia.arvi@kaist.ac.kr}

\author{Seoyoung Kim}
\orcid{0000-0002-0680-0856}
\affiliation{
  \institution{School of Computing \\ KAIST}
  \city{Daejeon}
  \country{Republic of Korea}
}
\email{youthskim@kaist.ac.kr}

\author{Juho Kim}
\orcid{0000-0001-6348-4127}
\affiliation{
  \institution{School of Computing \\ KAIST}
  \city{Daejeon}
  \country{Republic of Korea}
}
\affiliation{
  \institution{SkillBench}
  \city{Santa Barbara, CA}
  \country{USA}
}
\email{juhokim@kaist.ac.kr}

\renewcommand{\shortauthors}{Yeon Su Park et al.}

\begin{abstract}
    \input{sections/0_Abstract}
\end{abstract}

\begin{CCSXML}
<ccs2012>
   <concept>
       <concept_id>10003120.10003121.10011748</concept_id>
       <concept_desc>Human-centered computing~Empirical studies in HCI</concept_desc>
       <concept_significance>500</concept_significance>
       </concept>
 </ccs2012>
\end{CCSXML}

\ccsdesc[500]{Human-centered computing~Empirical studies in HCI}

\keywords{Self-Efficacy, Trust, Authorship, LLM Interaction Pattern, LLM-Assisted Writing, Human-AI Interaction}


\maketitle

\input{sections/1_Introduction}
\input{sections/2_Related_Work}
\input{sections/3_Method_Study}

\input{sections/4_Method_Analysis}

\input{sections/5_Result}
\input{sections/6_Discussion}
\input{sections/7_Limitation}
\input{sections/8_Conclusion}

\begin{acks}
This work was supported by the National Research Foundation of Korea (NRF) grant funded by the Korea government (MSIT) (No. RS-2025-00557726) and by Institute of Information \& Communications Technology Planning \& Evaluation (IITP) grant funded by the Korea government (MSIT) (No. RS-2024-00443251, Accurate and Safe Multimodal, Multilingual Personalized AI Tutors). We thank all of our study participants from Prolific for engaging positively in our study. We also thank all of the members of KIXLAB for their
helpful discussions and constructive feedback.
\end{acks}

\bibliographystyle{ACM-Reference-Format}
% \bibliography{main}
\bibliography{references}

\appendix
\input{sections/10_Appendix}

\end{document}

%% file: sections/0_Abstract.tex
Large language models (LLMs) are increasingly used as collaborative partners in writing. However, this raises a critical challenge of authorship, as users and models jointly shape text across interaction turns. Understanding authorship in this context requires examining users’ evolving internal states during collaboration, particularly self-efficacy and trust. Yet, the dynamics of these states and their associations with users’ prompting strategies and authorship outcomes remain underexplored. We examined these dynamics through a study of 302 participants in LLM-assisted writing, capturing interaction logs and turn-by-turn self-efficacy and trust ratings. Our analysis showed that collaboration generally decreased users’ self-efficacy while increasing trust. Participants who lost self-efficacy were more likely to ask the LLM to edit their work directly, whereas those who recovered self-efficacy requested more review and feedback. Furthermore, participants with stable self-efficacy showed higher actual and perceived authorship of the final text. Based on these findings, we propose design implications for understanding and supporting authorship in human-LLM collaboration.

%% file: sections/1_Introduction.tex
\section{Introduction} \label{intro}

Large language models (LLMs) have emerged as powerful general-purpose tools, leveraging their natural language capabilities and vast knowledge across diverse domains~\cite{brown2020language, zhao2023survey}. Hence, LLMs are now perceived as collaborative partners that co-produce content with users~\cite{lee2022coauthor}, extending their individual capabilities.

This collaborative potential is particularly visible in writing, a domain where ideas, structure, and phrasing are continuously negotiated and refined, making it a cognitively demanding and inherently iterative activity~\cite{sommers1980revision, flower1981cognitive}. Given this complexity, LLMs can play a complementary role in easing users' mental load and assisting their performance. Many previous studies have highlighted the benefits of such human-LLM collaborative writing, including enhanced productivity and confidence in writing~\cite{li2024value}, improved writing quality~\cite{dhillon2024shaping}, and facilitated idea exploration~\cite{wan2024felt}.

However, LLM-assisted writing introduces a critical challenge of authorship~\cite{huang2025authorship}. Here, users and LLMs continuously negotiate content across multiple turns. As LLMs increasingly handle generative and revisional roles that directly shape the text~\cite{mysore2025prototypical, guo2025pen}, boundaries between user and model contributions become less clear. Specifically, since writing is a subjective task where multiple solutions can be equally plausible, users may find it difficult to critically evaluate or distinguish the model’s input from their own. This blurring can lead to uncritical adoption of AI-generated ideas or phrasing~\cite{nguyen2024human}, as well as a weakened sense of agency and ownership of the final result~\cite{draxler2024ai}. This can further diminish users’ authorial voice and authenticity~\cite{hwang202580, guo2025pen} and raise concerns about the originality and integrity of the content they produce~\cite{chan2023ai}.

Understanding this authorship problem requires attending to users' evolving internal states during the writing process. Recent work has highlighted two complementary psychological constructs in human-AI collaboration: \textit{self-efficacy}---users’ belief in their ability to accomplish the work on their own---and \textit{trust}---their belief that the AI model will reliably support them~\cite{spitzer2025human, schemmer2023appropriate, chong2022human}. However, these studies often treat self-efficacy and trust as static attributes, failing to fully capture the dynamic nature of human-LLM collaborative writing. As users strategically prompt LLMs across multiple conversational turns, their perceptions of both themselves and the model can shift substantially~\cite{yang2023toward, nguyen2024human, chong2022human}. Without capturing these moment-to-moment changes and how they evolve over time, we cannot fully understand how authorship weakens, strengthens, or transforms throughout LLM-assisted writing.

To this end, we conducted an online experiment with 302 participants to examine the turn-by-turn dynamics of self-efficacy and trust in LLM-assisted writing. While self-efficacy and trust are traditionally conceptualized as stable psychological traits~\cite{bandura1977self, hoff2015trust}, we adopt a task-specific interpretation to capture them as dynamic states that fluctuate across turns. In our study, we define \textit{self-efficacy} as the user’s \textit{momentary} belief in their ability to complete this task on their own, and \textit{trust} as their \textit{momentary} belief that the LLM will reliably support this task. This view enables us to trace the dynamic trajectories of self-efficacy and trust throughout the writing process. Through this lens, we identify distinct patterns in how these constructs evolve, and examine how these patterns are manifested through users’ prompting strategies and are associated with their actual and perceived authorship of the produced text.

Our findings revealed distinct trajectory patterns: while self-efficacy tended to decrease over interaction turns, trust generally increased. We found that users with different self-efficacy trajectories exhibited distinct behavioral patterns. Participants with decreasing self-efficacy were more likely to ask the LLM for direct edits, often employing draft-to-edit and edit-to-edit prompt sequences, which resulted in reduced authorship. Participants who experienced initial self-efficacy drops but later recovered were more likely to request review and feedback from the LLM. In contrast, participants with stable self-efficacy reported higher actual and perceived authorship of the final text. Based on these insights, we outline design implications for LLM-assisted writing systems that can better understand and support user authorship during vulnerable moments.

The contributions of this paper are as follows:
\begin{itemize}
    \item Empirical characterization of turn-level trajectories of self-efficacy and trust in LLM-assisted writing.
    \item Findings on the relationship between self-efficacy trajectory patterns and (1) users' prompting strategies, and (2) their actual and perceived authorship.
    \item Design implications for understanding and supporting authorship in human-LLM collaboration.
\end{itemize}

%% file: sections/2_Related_Work.tex
\section{Related Work}
We review prior work across three domains to contextualize our investigation of human-LLM collaborative writing: (1) self-efficacy and trust in human-AI interaction, (2) user interaction patterns with AI systems, and (3) authorship in LLM-assisted writing.

\subsection{Self-Efficacy and Trust in Human-AI Interaction}
When users collaborate with LLMs across multiple turns, their decisions about when to rely on the model versus themselves depend on perceptions of their own capabilities and the system's. To examine these underlying psychological processes, we ground our work in two foundational constructs: \textit{self-efficacy} and \textit{trust}.

Bandura~\cite{bandura1977self,bandura1997self} defined self-efficacy as people's beliefs in their capability to organize and execute the actions required to accomplish a task. This construct has been shown to shape task choice, effort, and persistence across domains such as education and work~\cite{schunk1991self, stajkovic1998self}. In parallel, Mayer et al.~\cite{mayer1995integrative} conceptualized trust as a willingness to be vulnerable to another agent based on the expectation that the agent will act in ways that matter to the trustor. Subsequent work in automation and HCI adopts this framing to describe how users decide whether to rely on AI systems under uncertainty~\cite{ma2023should, zhang2020effect}. Together, self-efficacy and trust influence how people balance their own contributions with those of an AI system, shaping strategies for delegation, oversight, and joint performance~\cite{lubars2019ask, spitzer2025human, lee2004trust}.

A growing body of work in human–AI interaction has adopted these constructs, or closely related notions such as task-specific confidence, to explain how people engage with AI systems. Prior studies have examined how these beliefs shape selective adoption of AI suggestions~\cite{ma2023should}, prompting strategies~\cite{kumar2024guiding}, reliance calibration~\cite{he2023knowing,schemmer2023appropriate,li2025confidence}, and decision-making~\cite{buccinca2021trust,ma2024you,chong2022human}. In most research, however, self-efficacy and trust are treated as static with respect to a given task. They are typically measured once or twice (e.g., pre–post) or inferred from behavior aggregated over the entire task, and are assumed to remain relatively stable throughout interaction~\cite{kohn2021measurement, compeau1995computer}. As a result, moment-to-moment changes in these beliefs are rarely captured directly. While recent studies have begun to examine within-session adjustments in trust~\cite{wang2025viztrust, yang2023toward}, work that jointly tracks how users' beliefs about themselves and the AI co-evolve during complex, multi-turn collaboration remains limited.

In this work, we adopt a process-oriented perspective and examine self-efficacy and trust as dynamic, task-specific states that evolve over interaction. By modeling their turn-level trajectories, we provide a fine-grained account of how beliefs about oneself and the AI shape collaborative behavior in LLM-assisted writing.

\subsection{User Interaction Patterns with AI Systems}
Human-AI collaboration transformed traditional linear workflows into iterative, non-sequential ways where users move recursively between working stages~\cite{du2022understanding, reza2024abscribe}. To understand what makes a workflow effective, prior research primarily adopted an outcome-oriented approach~\cite{bansal2021does, guo2024decision}. For example, in education, measures often center on outcome quality or exam performance~\cite{kumar2024guiding, song2023enhancing, wang2024enhancing}. Creative tasks are often evaluated in terms of originality~\cite{wan2024felt, qin2025timing}, or diversity of output~\cite{qin2025timing, shaer2024ai}. In decision-making, success is typically measured through the decision accuracy~\cite{ma2023should, chong2022human, schemmer2023appropriate, buccinca2021trust}.

However, as AI systems become increasingly capable at producing high-quality outputs~\cite{zhao2023survey, brown2020language, kasneci2023chatgpt}, outcome-oriented evaluation becomes less informative for understanding human-AI interaction~\cite{lee2022evaluating, ibrahim2025towards}. When AI can reliably generate good results, the critical question shifts from ``what was produced'' to ``how did the collaboration unfold'' and ``what was the user's experience''~\cite{guo2025pen, lee2022evaluating}. In this context, examining users' interaction patterns becomes particularly important. Understanding these collaborative processes is essential because they reveal whether users maintain meaningful agency~\cite{guo2025pen} and cognitive engagement in the interaction~\cite{kosmyna2025your}, which directly impacts sustained learning~\cite{zhai2024effects} and long-term growth~\cite{kobiella2025efficiency}.

To better understand these collaborative processes, researchers have begun examining user interaction patterns through analysis of prompting strategies and conversational exchanges. For instance, McNichols et al.~\cite{mcnichols2025studychat} analyzed student-AI interactions to understand how different prompting strategies shape learning outcomes and engagement patterns. Similarly, Mysore et al.~\cite{mysore2025prototypical} examined in-the-wild interaction logs to identify prototypical human–AI collaboration behaviors, capturing recurring patterns in how users iteratively refine AI-generated text through follow-up prompts. Building on these approaches, our study investigates how users’ evolving internal states are reflected in their prompting strategies, in order to understand which prompting strategies and interaction patterns foster human agency in LLM-assisted writing.

\subsection{Authorship in LLM-Assisted Writing}
While previous research has shown that LLM-assisted writing improves authors' productivity~\cite{li2024value, dhillon2024shaping} and creative processes~\cite{wan2024felt, shaer2024ai, reza2024abscribe}, central to this collaboration is the problem of authorship~\cite{huang2025authorship, hwang202580, he2025contributions}. As LLMs increasingly function as coauthors~\cite{lee2022coauthor}, they blur the boundary between the contributions of humans and models, potentially diminishing human agency~\cite{qin2025timing, draxler2024ai} while complicating questions of ownership~\cite{hwang202580}. Hence, this raises fundamental questions about how to define, measure, and attribute authorship in the context of human-LLM collaborative writing.

To address this, prior works have predominantly relied on perceptual approaches that capture users' subjective assessments of their authorship contributions. Studies typically employ surveys asking participants to rate their sense of ownership~\cite{draxler2024ai, reza2025co}, creative self-efficacy~\cite{qin2025timing, mcguire2024establishing}, or perceived contribution~\cite{hwang202580}. Interview-based approaches have also been used to explore how writers negotiate boundaries between their work and AI-generated content~\cite{hwang202580, guo2025pen}. Some research has extended this approach by examining writers' attribution and responsibility in collaborative outputs~\cite{he2025contributions, reza2025co}.

However, these perceptual approaches have some limitations. Studies examining varying degrees of human versus AI contribution reveal differences between users' subjective authorship claims and their objective contribution levels~\cite{draxler2024ai, he2025contributions}, yet few studies have systematically examined both perceptual and objective dimensions together. In this work, we address this gap by jointly analyzing perceptual authorship and objective AI output utilization behaviors, establishing a more comprehensive measure of \textit{actual} and \textit{perceived} authorship in the context of LLM-assisted writing.

%% file: sections/3_Method_Study.tex
\section{Method}
We investigate the dynamics of users' self-efficacy and trust in the context of LLM-assisted argumentative essay writing. As defined earlier (Section~\ref{intro}), we conceptualize \textit{self-efficacy} as the writer's momentary belief in their ability to complete the given task independently, and \textit{trust} as their momentary belief that the LLM will reliably support the task. Our research questions are as follows:

\begin{itemize}
    \item RQ1. What trajectory patterns emerge in users' self-efficacy and trust, and how do they interact over time?
    \item RQ2. How are users’ prompting strategies associated with self-efficacy and trust trajectory patterns?
    \item RQ3. How are actual and perceived authorship associated with self-efficacy and trust trajectory patterns?
\end{itemize}

In this study, we focus on the trajectory patterns of users' self-efficacy and trust rather than their static scores for two primary reasons. First, self-efficacy and trust are highly subjective measures that vary across individuals as each person has a unique personal standard and interpretation of the scale~\cite{van2002theory, xu2024examining}. By examining within-person changes over time instead of a single score, we control for these individual differences. Second, LLM-assisted writing is a dynamic, iterative process where user perceptions evolve through interaction~\cite{wan2024felt, flower1980cognition, siddiqui2025script}. Trajectories capture this progressive aspect and allow us to examine how different prompting strategies and authorship experiences are associated with them. 

In this section, we explain the details on how we designed and conducted our study and analyzed the data collected. The study was reviewed and approved by the Institutional Review Board (IRB) of our institution prior to conducting the study.

\subsection{Study Design}
Here, we outline how we conducted our study. 

\subsubsection{Participants}
We aimed to recruit a minimum of 300 participants after applying our exclusion criteria. This target was established following the criteria of VanVoorhis and Morgan~\cite{vanvoorhis2007understanding}, which recommend around 30 observations per cell to achieve 80\% power in group comparisons. Our pilot study ($N=20$) indicated the least frequent trajectory pattern occurred in about 10\% of the sample, thus a sample of 300 would provide approximately 30 participants in the smallest group, meeting power requirements.

We recruited a total of 313 initial participants through Prolific\footnote{\url{https://www.prolific.com/}}. Since the task of our study was to write an argumentative essay in English, we used the preliminary filtering function of Prolific to only recruit native English speakers, ensuring consistent language proficiency across participants. We also restricted recruitment to participants with approval ratings of 95\% or higher, which indicates high-quality participation in previous studies.

To observe participants' natural strategies of LLM use during the writing task, we recruited individuals with recent experience. A custom screening survey was used to confirm they had used an LLM within the last month. This qualification ensured that participants' use of the tool was based on preference or strategy, rather than a lack of familiarity. Participants who did not meet this criterion were screened out and received a nominal fee of 0.1 GBP. 

Following data collection, we excluded 11 participants (3.5\%) from analysis as a quality control. One participant who took over two hours was excluded due to clear disengagement from the task. The remaining 10 participants spent less than five minutes on writing. We determined this duration insufficient to read the writing prompt, wait for AI responses to generate, review the output, and compose a response. This threshold fell below the 5th percentile (7.00 minutes) of writing durations, representing an inadequate timeframe for meaningful task completion. Beyond time-based criteria, our quality control also included reviewing participants' open-ended explanations for their ratings and post-survey responses (see Section~\ref{method:study}) to identify off-topic or placeholder answers. However, no additional participants met these exclusion criteria. 

After applying exclusion criteria, our final data consisted of 302 participants. We paid all eligible participants ($N=313$) 6.75 GBP ($\approx$ 9.11 USD), regardless of whether they were included in the final sample or not. The median study completion time was 39.21 minutes, with the median writing time of 22.97 minutes.

To motivate participants to put their best effort into the writing task, we offered a performance-based incentive. Participants were informed that the top 10\% of submissions would receive an additional 6.75 GBP based on essay quality, doubling their total payment. The submitted essays were graded using the College Board's scoring rubric and guidelines\footnote{\url{https://apcentral.collegeboard.org/courses/ap-english-language-and-composition}\label{fn:cb}}. To support consistency in evaluation, we generated initial scores with the LLM, and one author reviewed such scores alongside their brief justifications to adjust them to determine final grades. Overall, the author adjusted a total of 6.10\% of the LLM-generated scores, and 34 participants (11.26\%) received the additional bonus payment. We outline the detailed procedure as well as the prompt used in the Appendix~\ref{appendix:grading}.

\subsubsection{Task} \label{method:task}
Participants were asked to complete an argumentative essay as the main task. The essay prompt was drawn from the College Board AP English Language and Composition exam\hyperref[fn:cb]{\textsuperscript{\ref*{fn:cb}}}, similar to the setting of Siddiqui et al.~\cite{siddiqui2025ai}. We chose this exam for being a validated assessment with its standardized evaluation criteria, clear scoring guidelines, and accessibility for a general audience.

Participants were asked to write a minimum of 300 words within a 30-minute time limit, adapted from the original exam setting. During the task, they could freely interact with the embedded LLM through the study interface (see Section~\ref{method:interface}) to support different aspects of writing, such as idea generation, drafting, and revision.

\subsubsection{Study Procedure} \label{method:study}
The study consisted of three main phases: (1) pre-survey, (2) writing task, and (3) post-survey.

Firstly, participants completed a brief pre-survey measuring their initial, baseline self-efficacy and trust levels. They responded to two questions on 7-point Likert scales (1 = Not at all, 7 = Completely):

\begin{itemize}
\item \textbf{[Self-Efficacy]} How confident are you in your ability to complete academic English writing tasks on your own?
\item \textbf{[Trust]} How much do you trust LLMs to support you in completing academic English writing tasks?
\end{itemize}

These baseline ratings served dual purposes: establishing initial self-efficacy and trust scores for subsequent analysis, and providing participants with reference anchors for the in-the-moment ratings they would provide during the main task. 

Secondly, participants proceeded to the main writing task as described in Section~\ref{method:task}. To capture the dynamics of self-efficacy and trust, we collected in-the-moment ratings at the end of each interaction turn. Before submitting a new prompt, participants were required to rate their current levels of self-efficacy and trust on a 7-point Likert scale (1 = Not at all, 7 = Completely):

\begin{itemize}
\item \textbf{[Self-Efficacy]} At this point, how confident are you in your ability to complete this writing task on your own?
\item \textbf{[Trust]} At this point, how much do you trust the LLM to support you in completing this writing task?
\end{itemize}

Note that the same parallel questions from the pre-survey were repeatedly asked for each turn, allowing us to track their dynamic changes and treat both constructs as states that evolve throughout the interaction. The timing was chosen to allow participants sufficient opportunity to process the LLM's response and incorporate it into their draft, enabling more informed and reflective ratings.

The task concluded either when a participant met the 300-word requirement and chose to proceed, or automatically after 30 minutes had passed. In either case, participants had to complete the ratings for their final interaction turn before moving on. Afterward, participants viewed a chronological timeline of their LLM interactions and provided brief open-ended explanations for each self-efficacy and trust rating they had given during the task.

Lastly, participants finished the task by completing a post-survey. They rated their perceived authorship of their final essay using two dimensions---ownership and agency---each measured on a 7-point Likert scale (1 = Not at all, 7 = Completely):  

\begin{itemize}
\item \textbf{[Ownership]} How much do you feel like you are the author of the resulting text?
\item \textbf{[Agency]} I felt like I was in control of the writing process during the task.
\end{itemize}

These questions were adapted from Qin et al.~\cite{qin2025timing} and Draxler et al.~\cite{draxler2024ai}, respectively. In the post-survey, participants were also asked whether responding to in-the-moment ratings had influenced their interaction with the LLM, and, if so, in what ways.

We designed the rating questions to be as simple as possible to maintain natural interaction while still capturing the dynamics of self-efficacy and trust throughout the writing process. In the last post-survey question, 76.7\% of participants responded that the rating process did not affect their interactions. However, 19.2\% noted that it was either distracting or made them more mindful about their choices. We acknowledge this limitation in Section~\ref{limitation}.

\subsubsection{Study Interface} \label{method:interface}
\begin{figure*}[hbt]
  \centering
  \newcommand{\colw}{0.50\linewidth}
  \makebox[\linewidth][c]{%
    \begin{subfigure}[t]{\colw}
      \includegraphics[width=\textwidth]{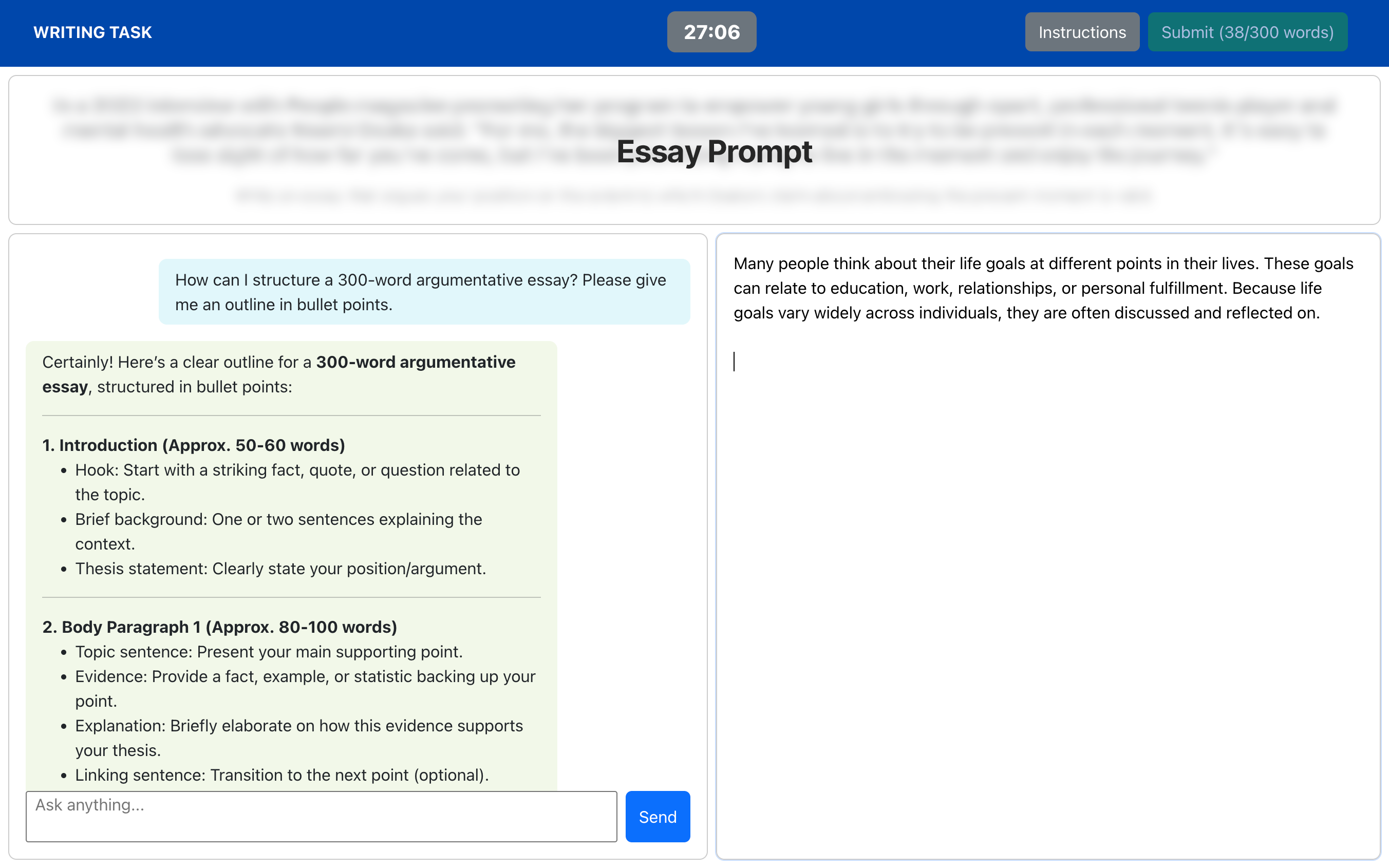}
      \caption{Writing Interface}\label{fig:platform1}
      \Description{Screenshot of the study platform interface. The essay prompt is shown at the top, the LLM chatbot area is on the left, and the participant’s essay writing area is on the right.}
    \end{subfigure}
    \begin{subfigure}[t]{\colw}
      \includegraphics[width=\textwidth]{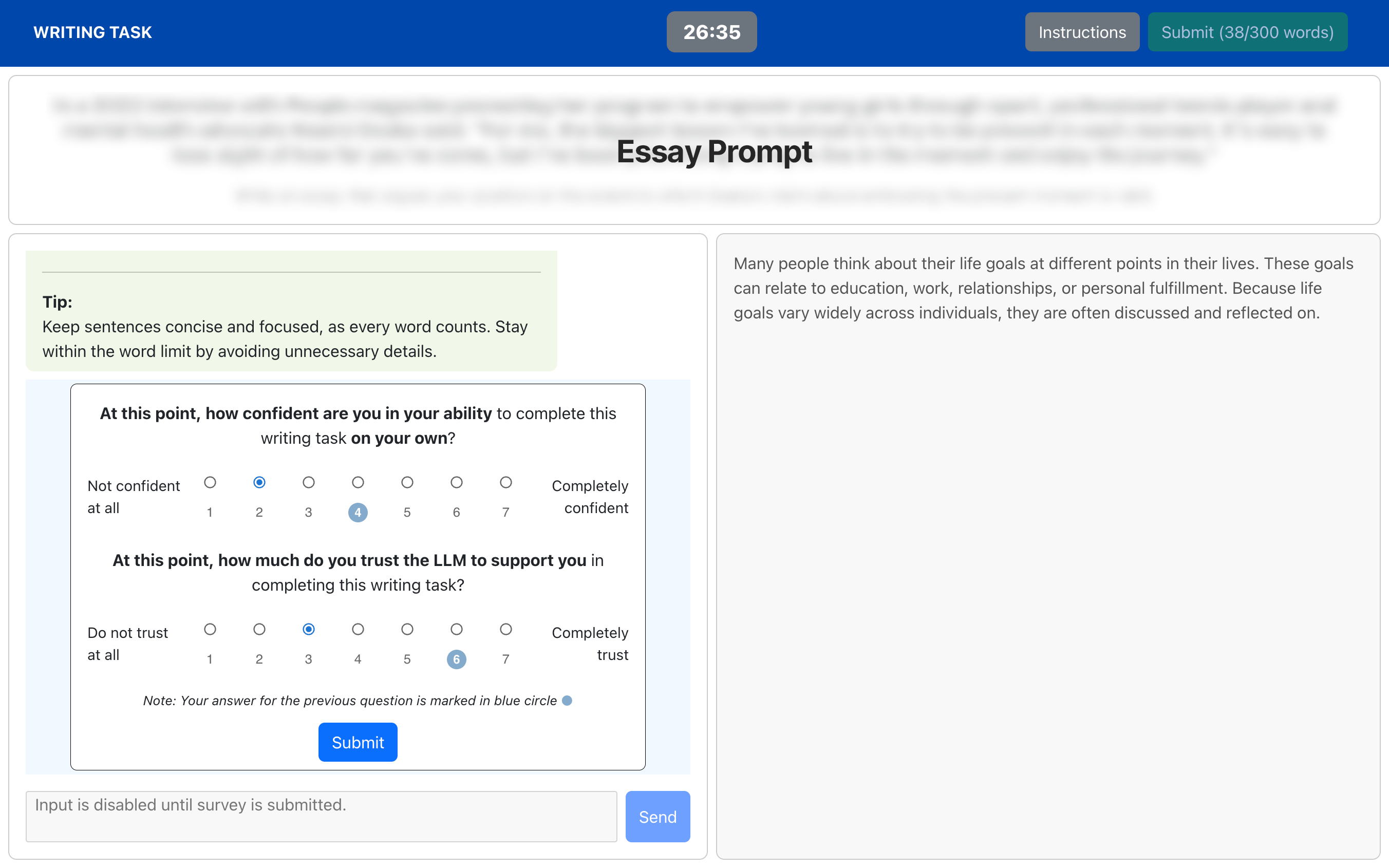}
      \caption{Self-Efficacy \& Trust Ratings}\label{fig:platform2}
      \Description{Screenshot of the self-efficacy and trust rating interface, shown within the LLM chatbot area. Participants respond on two 7-point scales: one for their confidence in completing the task and one for their trust in the LLM’s support.}
    \end{subfigure}%
  }
  \caption{Overview of the study interface: (a) LLM-assisted writing workspace and (b) real-time self-efficacy and trust rating}
  \Description{Overview of the study platform showing two interfaces: (a) the writing workspace with the LLM chatbot and essay area, and (b) the rating interface for self-efficacy and trust.}
  \label{fig:platform}
\end{figure*}

We deployed a custom web interface to conduct the study on Prolific, building on the design of Li et al.~\cite{li2024value} (Figure~\ref{fig:platform}). As shown in Figure~\ref{fig:platform1}, the interface consisted of two primary panels: a chat panel on the left for interacting with the LLM, and a text editor on the right where participants could draft their essay. The chat panel was implemented using vanilla \texttt{gpt-4.1}\footnote{\url{https://platform.openai.com/docs/models/gpt-4.1}} to replicate a representative real-world usage environment. 

As explained in Section~\ref{method:study}, the interface required users to rate their in-the-moment self-efficacy and trust before sending the next prompt, as shown in Figure~\ref{fig:platform2}. For each interaction turn with LLM, the system recorded the user's prompt, the LLM's response, the corresponding self-efficacy and trust ratings, the state of the essay at that moment, with timestamps at each logged action.

To facilitate this rating process, the interface had two features. First, to help participants anchor their current feelings, a visual cue indicating their rating from the previous turn was displayed for each question. Second, to encourage deliberate reflection on their momentary self-efficacy and trust, both the prompt input and essay editor were disabled until the ratings were submitted.

Furthermore, to ensure data integrity and prevent gaming behaviors, the interface strictly prohibited URL manipulation to skip or revisit study phases and automatically saved participants’ progress so that page refreshes would not affect task flow or recorded data. All refresh events were logged, and the authors manually reviewed these cases and confirmed that no participants displaying such noncompliant behaviors remained in the final data.

%% file: sections/4_Method_Analysis.tex
\subsection{Data Analysis}
We explain how we analyzed the collected data to answer each RQ. Our analyses consist of five main components: (1) identifying trajectory patterns of self-efficacy and trust (RQ1--3), (2) modeling their dynamic interactions using mixed-effects models (RQ1), (3) qualitatively coding user prompts to capture underlying intentions (RQ2), (4) analyzing prompting strategies through distribution of prompt intentions and their transition dynamics (RQ2), and (5) measuring actual authorship by comparing user writing with LLM-generated content (RQ3). Other statistical comparisons are conducted as needed and reported in the results directly (Section ~\ref{result}).

\subsubsection{Identifying Trajectory Patterns of Self-Efficacy and Trust} \label{method:pattern}
A central component of our analysis for all research questions is identifying trajectory patterns of self-efficacy and trust dynamics (RQ1--3). Here, we describe our analysis approach to classifying each participant's trajectories into distinct patterns.

To categorize these trajectories, we first established a statistically meaningful threshold for distinguishing reliable change from random measurement noise. Following prior works that similarly employed repeated administration of rating scales to assess within-person change in psychological constructs~\cite{parabiaghi2014defining, de2018treatment}, we adopted the Minimal Detectable Change (MDC) threshold~\cite{haley2006interpreting, jacobson1992clinical} to distinguish reliable change from measurement error. At a 95\% confidence interval, we found the theoretical threshold to be 2.080 points for self-efficacy and 2.037 points for trust. Given the discrete integer nature of the 7-point Likert scale, rounding the MDC up to 3 points would set a threshold that is disproportionately large relative to the scale’s resolution, making it less sensitive to meaningful changes. Therefore, we chose the nearest integer, 2 points, as the practical MDC cutoff indicating a significant change.

Consistent with established MDC-based practices to classify individuals into distinct groups~\cite{jacobson1992clinical, abrahamse2012parent, de2018treatment}, we used this threshold to first identify the initial trajectory patterns based on net change between final and initial scores:

\begin{itemize}
    \item \textbf{Increase}: The final score was $\geq 2$ points higher than the initial score.
    \item \textbf{Decrease}: The final score was $\geq 2$ points lower than the initial score.
    \item \textbf{Stable}: All other cases where the final score was within $\pm 1$ points of the initial score.
\end{itemize}

However, simple pre-post comparisons obscure meaningful within-session dynamics, as participants who appear stable in terms of net change may nevertheless exhibit qualitatively different intermediate trajectories during the interaction. To capture these dynamics, we further decomposed the \textit{stable} pattern into three categories by examining whether any intermediate ratings crossed the same MDC-based 2-point threshold during the interaction:

\begin{itemize}
    \item \textbf{Recovery}: The final score dropped $\geq 2$ points but returned to within $\pm 1$ point of initial.
    \item \textbf{Reversion}: The final score increased $\geq 2$ points but returned to within $\pm 1$ point of initial.
    \item \textbf{Stable}: All other cases which did not meet recovery or reversion criteria.
\end{itemize}

Through this process, we established five distinct trajectory patterns: \textit{increase}, \textit{decrease}, \textit{stable}, \textit{recovery}, and \textit{reversion}. To maintain adequate power, we only included the patterns present in $\ge 10\%$ of participants in the main analyses.

\renewcommand{\arraystretch}{1.25}
\begin{table*}[t!]
\caption{Intention Categories and Definitions of User Prompts}
\label{tab:tagging}
\begin{tabular}{c|l|l|l}
\toprule
\textbf{Category} & \multicolumn{1}{c|}{\textbf{Subcategory}} & \multicolumn{1}{c|}{\textbf{Description}} & \multicolumn{1}{c}{\textbf{Example}} \\ \hline
\multirow{3}{*}{\textbf{Drafting}} & Full Draft & \begin{tabular}[c]{@{}l@{}}User requests a complete essay \end{tabular} & \begin{tabular}[c]{@{}l@{}}[\textit{Essay Prompt}]\\ Please write a first draft.\end{tabular} \\ \cline{2-4} 
 & Partial Draft & \begin{tabular}[c]{@{}l@{}}User requests specific essay sections \end{tabular} & \begin{tabular}[c]{@{}l@{}}Can you please help me come up \\ with a conclusion? \end{tabular} \\ \cline{2-4} 
 & Outline & \begin{tabular}[c]{@{}l@{}}User requests the overall structure \\ or organization of an essay \end{tabular} & \begin{tabular}[c]{@{}l@{}}Can you help me make a short \\ outline for my essay?\end{tabular} \\ \hline
\textbf{Editing} & \multicolumn{1}{c|}{---} & \begin{tabular}[c]{@{}l@{}}User requests direct edits or \\ guidance to improve their writing. \end{tabular} & \begin{tabular}[c]{@{}l@{}}Can you refine this and polish it up? \\ This is my essay so far. [\textit{Essay Draft}]\end{tabular} \\ \hline
\textbf{Ideating} & \multicolumn{1}{c|}{---} & \begin{tabular}[c]{@{}l@{}}User seeks ideas, arguments, \\ or exploratory input.\end{tabular} & \begin{tabular}[c]{@{}l@{}}What are the pros and cons \\ of [\textit{Person Name}]'s statement?\end{tabular} \\ \hline
\multirow{3}{*}{\textbf{\begin{tabular}[c]{@{}c@{}}Information\\ Searching\end{tabular}}} & Context & \begin{tabular}[c]{@{}l@{}}User requests context information \\ relevant to the essay topic.\end{tabular} & \begin{tabular}[c]{@{}l@{}}Give me a brief background \\ check on [\textit{Person Name}].\end{tabular} \\ \cline{2-4} 
 & Evidence & \begin{tabular}[c]{@{}l@{}}User requests specific evidence \\ to support an argument. \end{tabular} & \begin{tabular}[c]{@{}l@{}}Can you find a scientific study \\ that shows the [\textit{Argument}]?\end{tabular} \\ \cline{2-4} 
 & Writing & \begin{tabular}[c]{@{}l@{}}User seeks general information \\ about writing practices.\end{tabular} & \begin{tabular}[c]{@{}l@{}}What is the best way to structure \\ an argumentative essay?\end{tabular} \\ \hline
\textbf{Reviewing} & \multicolumn{1}{c|}{---} & \begin{tabular}[c]{@{}l@{}}User requests feedback or \\ evaluation on their writing. \end{tabular} & \begin{tabular}[c]{@{}l@{}}This is what I have so far. What \\ do you think? {[\textit{Essay Draft}]}\end{tabular} \\ \hline
\multirow{3}{*}{\textbf{Others}} & Language & \begin{tabular}[c]{@{}l@{}}User seeks assistance with \\ spelling, grammar, or wording\end{tabular} & Synonym for savor? \\ \cline{2-4} 
 & No Intent & \begin{tabular}[c]{@{}l@{}}User provides content as a \\ follow-up, without a new intent\end{tabular} & [\textit{Essay Draft}] \\ \cline{2-4} 
 & Others & \begin{tabular}[c]{@{}l@{}}User engages in conversation \\ unrelated to essay writing\end{tabular} & Thanks. I'm happy with this now.
\\ \bottomrule
\end{tabular}
\Description{Table of intention categories of user prompts in the study. The table lists categories such as drafting, editing, ideating, information searching, reviewing, and others, with subcategories, descriptions, and example prompts illustrating each type.}
\end{table*}

\subsubsection{Modeling the Dynamics of Self-Efficacy and Trust}
\label{method:mixed-effects}
To investigate how self-efficacy and trust interact over time (RQ1), we analyzed their bidirectional relationship using a linear mixed-effects model. This was chosen as our data contained repeated measures from each participant across multiple interaction turns, violating the independence assumption of standard regression models.

Our primary analysis involved fitting two complementary models to examine the relationship: (1) one predicting \textit{self-efficacy} scores from \textit{turn}, \textit{trust}, and their interaction (\textit{turn} $\times$ \textit{trust}), and (2) another one predicting \textit{trust} scores from \textit{turn}, \textit{self-efficacy}, and their interaction (\textit{turn} $\times$ \textit{self-efficacy}). Here, \textit{turn} was included as a key predictor in both models to capture how the relationship between self-efficacy and trust evolved over the course of the interaction.

We mean-centered all continuous predictors so that zero corresponded to the sample mean. Under this specification, each main effect represents the influence of a predictor when the interacting variable is fixed at its average level.

We tested two alternative random-effects structures. The first only included random intercepts, while the second included both random intercepts and random slopes for \textit{turn}. Random intercepts account for individual baseline differences, while random slopes account for individual differences in the rate of change over time. A likelihood ratio test showed that the model with random slopes had a significantly better fit, so we report results from this specification.

\subsubsection{Qualitative Analysis of User Prompts}
\label{method:prompt}
To examine users’ prompting strategies (RQ2), we qualitatively coded the underlying intentions of user prompts from the interaction logs. Each prompt was paired with its corresponding LLM response and ordered chronologically per participant to preserve contextual information.

The process was divided into three main steps. First, two authors independently reviewed 5.5\% of the user prompts to identify recurring intentions. Here, LLM responses were visible during this step to ensure that prompts referring to prior outputs could be properly interpreted. Inter-rater reliability was high in this round ($\kappa=0.867$). Any discrepancies or uncertainties were discussed, and a set of initial intention categories was established. 

Second, the authors applied these initial categories to an additional 10.1\% of the data to validate their stability and agreement. No changes to the categories were required, and inter-rater reliability remained substantially high ($\kappa=0.789$). 

Lastly, using the finalized categories, the remaining prompts were divided between the two authors and coded separately. Prompts could have multiple intentions when applicable. This process resulted in five main intention categories: \textit{drafting}, \textit{editing}, \textit{ideating}, \textit{information searching}, and \textit{reviewing}, which are described in Table~\ref{tab:tagging}.

\subsubsection{Analyzing Prompting Strategies}
\label{method:strategy}
To understand user prompting strategies more holistically (RQ2), we performed two analyses based on the categorized prompt intentions. These analyses focused on investigating both the overall distribution of individual prompt intentions and the dynamic patterns of prompt transitions.

First, we analyzed the proportional usage of each prompting intention. Since participants completed different numbers of interaction turns, we normalized them by computing proportions relative to each person's total number of prompts. This measure allowed us to compare which intentions were strategically preferred across different trajectory patterns.

Second, to capture the dynamic and interactive nature, we analyzed transitions between consecutive prompting strategies. A static analysis of individual prompt proportions would not fully reveal how users' strategies evolved over the course of an interaction. We created a sequence of consecutive prompt pairs for each user (e.g., from prompt $P_i$ to $P_{i+1}$). We then calculated the proportion of each transition type (e.g., from \textit{drafting} to \textit{editing}) relative to the total number of transitions. This analysis provided insight into the preferred strategic sequences of each user group, revealing patterns that individual prompting preferences cannot capture on their own.

Since there are a large number of possible transitions with five prompt categories, we only included those transitions exhibited by at least 10\% of participants in our main analysis to maintain statistical power and robustness.

\subsubsection{Measuring Actual Authorship}
\label{method:authorship}
To analyze actual and perceived authorship (RQ3), we measured how participants incorporated LLM-generated content into their essays. While perceived authorship was captured through post-survey, actual authorship required an objective measurement of content overlap between LLM responses and user writings.

For measuring actual authorship, we employed two complementary approaches to capture different aspects of content incorporation: (1) lexical overlap and (2) semantic similarity. This combination was useful as direct copying (lexical overlap) and rephrasing (semantic similarity) are both common and distinct ways users integrate LLM-generated content into their work~\cite{chanpradit2024english, chan2023ai, wang2024enhancing}. Lexical overlap was measured using ROUGE-3, which calculates recall-based tri-gram overlap by measuring how much of the reference text (LLM response) appears in the candidate text (user's writing). Semantic similarity was measured via text embeddings using \texttt{gemini-embedding-001}\footnote{\url{https://ai.google.dev/gemini-api/docs/embeddings}}. We captured conceptual overlap by computing the cosine similarity of high-dimensional vector embeddings from both the LLM response and user text.

We examined content adoption by comparing LLM-generated text across the entire interaction with the final essay. For each LLM response, we collected all preceding user messages to identify content the user had already contributed. We then extracted only the novel content which are originally generated by the LLM, excluding portions that restated user input since LLMs frequently rephrase user ideas. This LLM-original content and participants' final essays were tokenized and compared using both lexical and semantic similarity measures. This approach captures cases where users selectively adopted LLM content from various interaction points, regardless of when the adoption occurred.

%% file: sections/5_Result.tex
\section{Result} \label{result}
In this section, we first present descriptive statistics as an overview and discuss each RQ in detail. We present our study results on users' self-efficacy and trust patterns and changes over time (RQ1), their association with prompting strategies (RQ2), and their association with actual and perceived authorship of the outcome (RQ3).

\subsection{Descriptive Statistics}
Our study collected a total of 1,410 user prompts and LLM response pairs. Each participant made an average of 4.67 turns ($SD=3.87$, $min=1.0$, $max=25.0$), showing substantial variation and diversity in interaction patterns. Specifically, 368 prompts were labeled as \textit{drafting} (26.1\%), 305 as \textit{editing} (21.6\%), 185 as \textit{ideating} (13.1\%), 230 as \textit{information searching} (16.3\%), and 105 as \textit{reviewing} (7.4\%), with the remaining 217 prompts categorized as \textit{others} (15.4\%).

Participants' self-efficacy scores on a 7-point Likert scale had a mean of 5.07 ($SD=1.74$, $min=1$, $max=7$), indicating moderate self-efficacy levels. The average change in self-efficacy per turn was -0.16 ($SD=1.01$, $min=-6$, $max=5$), suggesting that participants' self-efficacy decreased slightly with each interaction turn.

On the other hand, participants' trust levels in the LLM were higher than that of self-efficacy, with a mean of 6.00 ($SD=1.28$, $min=1$, $max=7$). The average change in trust per turn was 0.14 ($SD=0.72$, $min=-5$, $max=4$), indicating that participants' trust in the LLM increased slightly with each interaction turn.

\subsection{RQ1. What trajectory patterns emerge in users' self-efficacy and trust, and how do they interact over time?} \label{RQ1}

We first characterize the main trajectory patterns of users' self-efficacy and trust during an interaction session with the LLM. Then, we examine how these two constructs dynamically interact over turns using linear mixed-effects regressions.

\subsubsection{Trajectory Patterns of Self-Efficacy \& Trust} \label{RQ1_1}

\begin{figure*}[t!]
  \centering
  \newcommand{\colw}{0.33\linewidth}
  \begin{subfigure}[t]{\colw}
    \includegraphics[width=\textwidth]{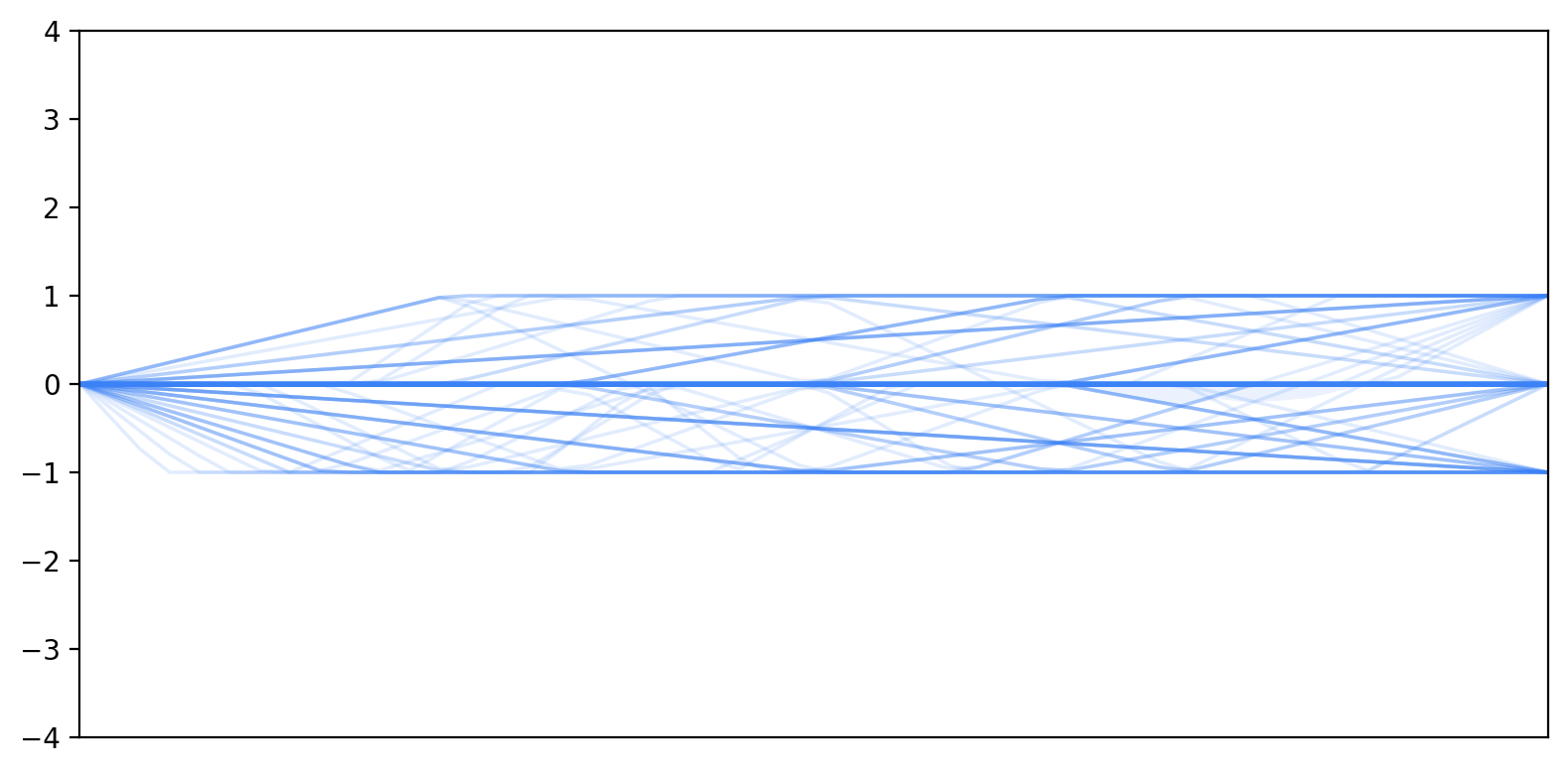}
    \caption{Self-Efficacy — Stable ($N=171$)}\label{fig:conf-stable}
    \Description{Line plot showing participants’ self-efficacy scores remaining relatively constant over time, with the median line flat and a narrow interquartile range.}
  \end{subfigure}
  \begin{subfigure}[t]{\colw}
    \includegraphics[width=\textwidth]{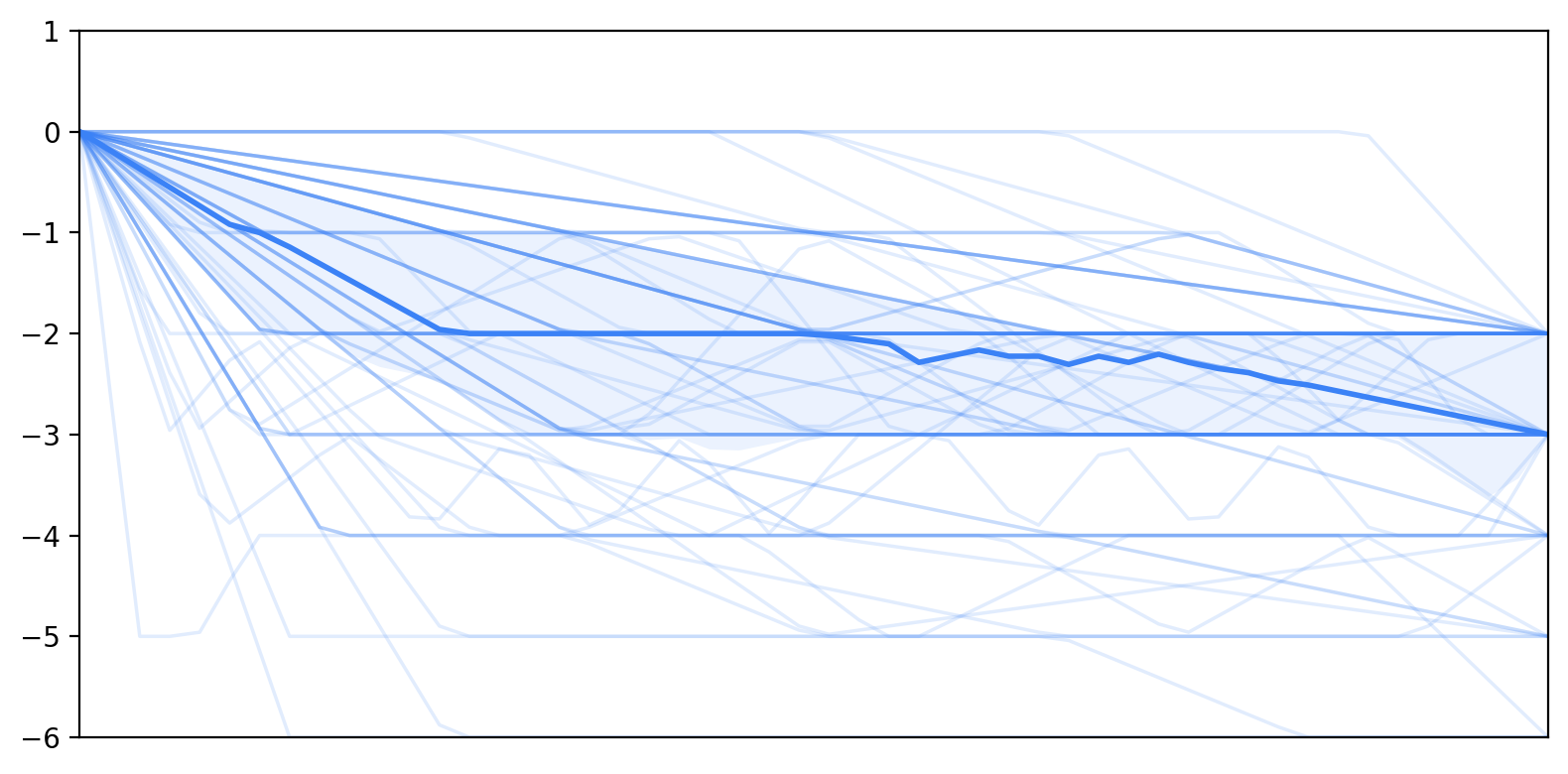}
    \caption{Self-Efficacy — Decrease ($N=83$)}\label{fig:conf-decrease}    
    \Description{Line plot showing a downward trend in participants’ self-efficacy scores over time, with the median decreasing steadily.}
  \end{subfigure}
  \begin{subfigure}[t]{\colw}
    \includegraphics[width=\textwidth]{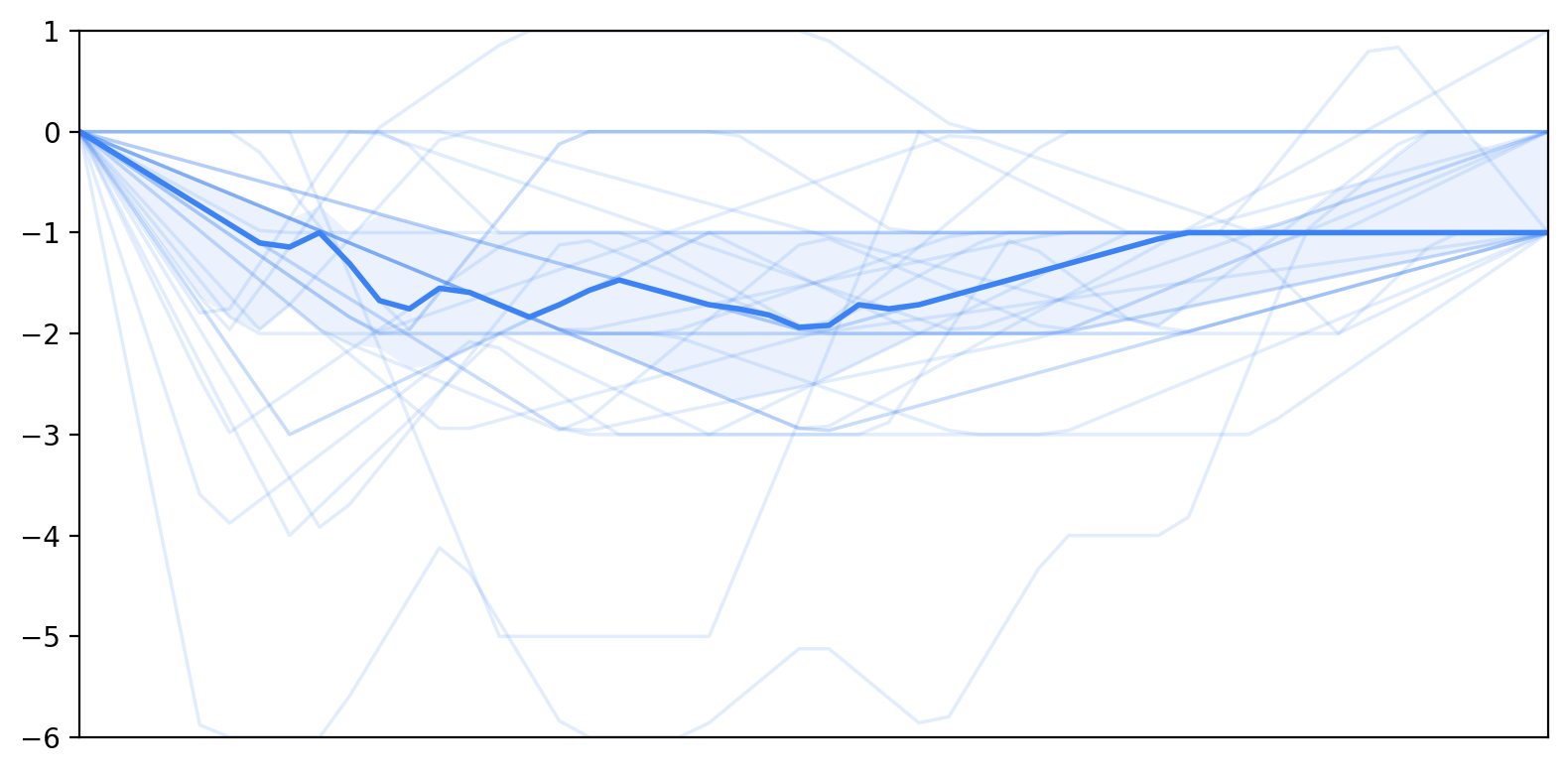}
    \caption{Self-Efficacy — Recovery ($N=33$)}\label{fig:conf-recovery}   
    \Description{Line plot showing participants’ self-efficacy scores initially declining but then rising again toward baseline, with a U-shaped median trend.}
  \end{subfigure}
    \par\vspace{0.3\baselineskip}
  \begin{subfigure}[t]{\colw}
    \includegraphics[width=\textwidth]{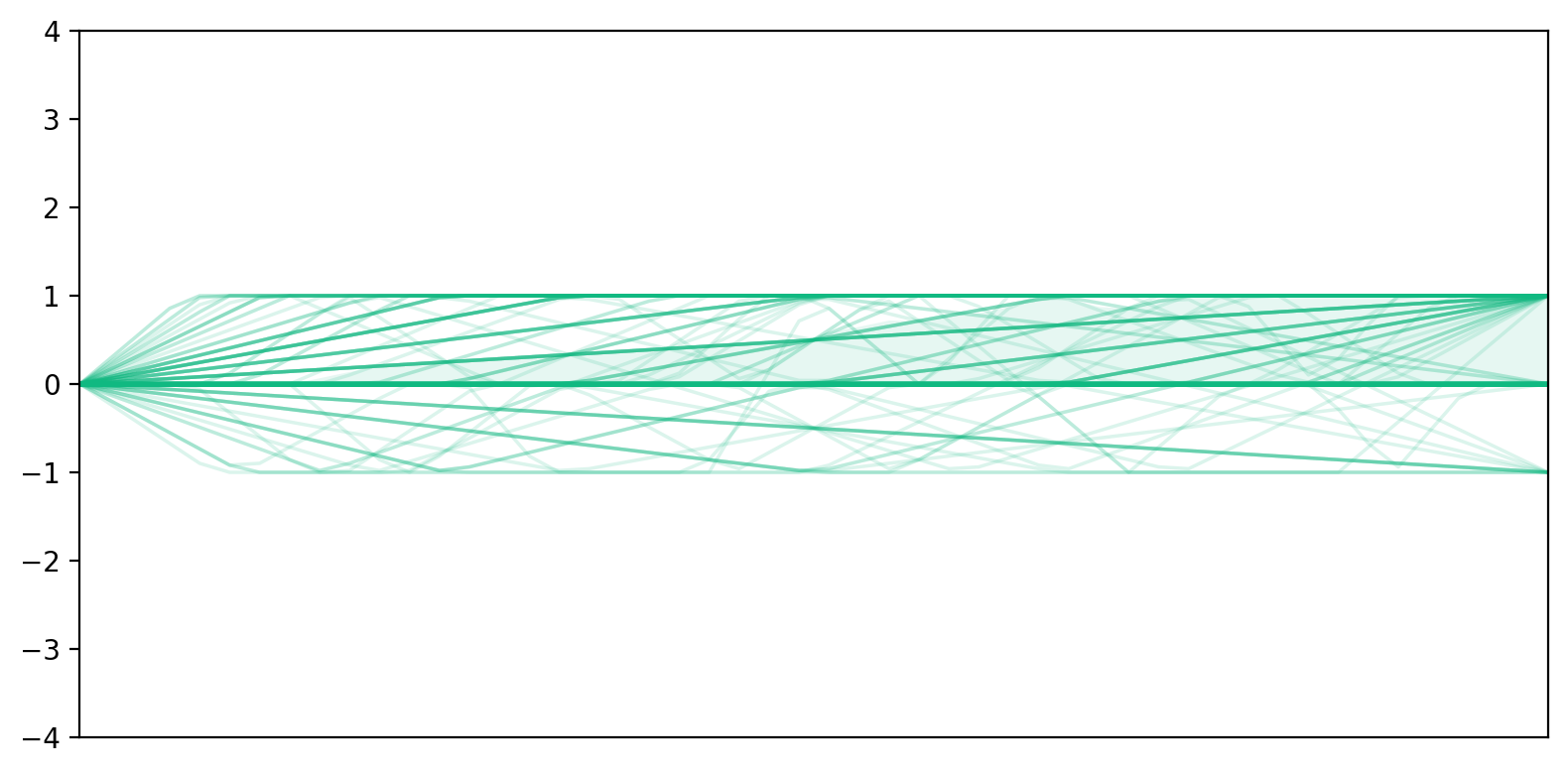}
    \caption{Trust — Stable ($N=231$)}\label{fig:trust-stable}    
    \Description{Line plot showing participants’ trust scores staying steady over time, with the median line flat.}
  \end{subfigure}
  \begin{subfigure}[t]{\colw}
    \includegraphics[width=\textwidth]{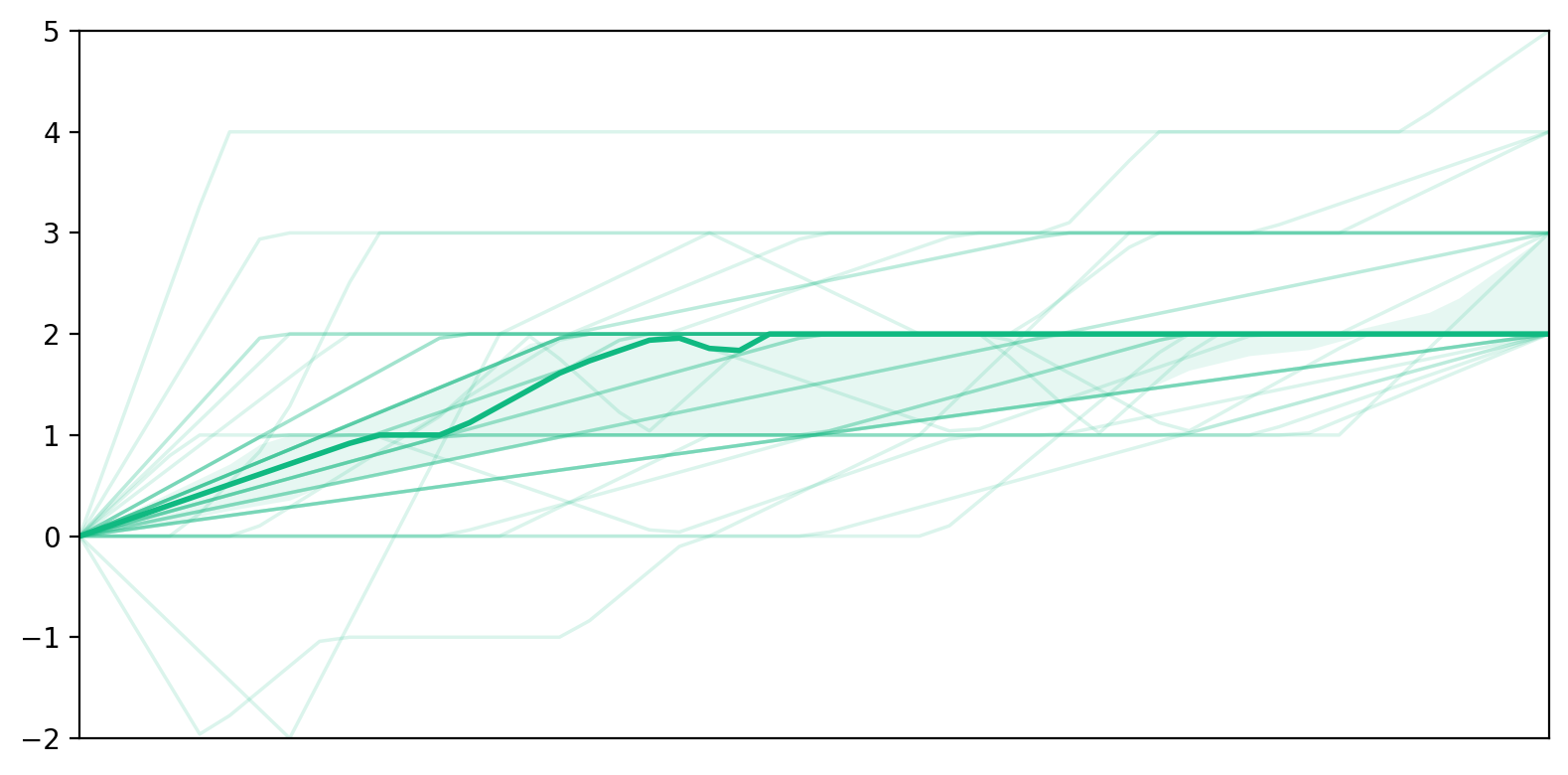}
    \caption{Trust — Increase ($N=47$)}\label{fig:trust-increase}   
    \Description{Line plot showing participants’ trust scores rising over time, with the median line trending upward.}
  \end{subfigure}
  \begin{subfigure}[t]{\colw}
    \mbox{}
  \end{subfigure}
  \caption{Normalized overlays of self-efficacy and trust trajectories for the five major trajectory patterns. The horizontal axis shows the time-normalized turn-by-turn progression from initial pre-survey rating (left) to final rating (right). The vertical axis shows the relative change in Likert-scale points after zero-centering all trajectories at each participant's pre-survey rating. Thin translucent lines represent individual participants, the solid line shows the group median, and the shaded band indicates the interquartile range.}
  \Description{Five representative trajectory patterns of self-efficacy and trust observed across participants: stable self-efficacy, decreasing self-efficacy, recovering self-efficacy, stable trust, and increasing trust. Each plot shows individual traces, the median, and the interquartile range.}
  \label{fig:pattern_overlay}
\end{figure*}

We categorized users’ self-efficacy and trust into five trajectory patterns as explained in Section ~\ref{method:pattern}. We first report the distribution of participants across patterns and highlight primary patterns for self-efficacy and trust to focus on later analyses, illustrated in Figure~\ref{fig:pattern_overlay}.

For self-efficacy, among 302 participants of our study, 56.6\% ($N=171$) were classified as \textit{stable}, 27.5\% ($N=83$) as \textit{decrease}, 10.9\% ($N=33$) as \textit{recovery}, 4.0\% ($N=12$) as \textit{increase}, and 1.0\% ($N=3$) as \textit{reversion}. Overall, net decreases were more frequent than net increases. Hence, \textbf{we focus on three major trajectory patterns for our subsequent analyses of self-efficacy:} \textbf{\textit{stable}} (Fig~\ref{fig:conf-stable}), \textbf{\textit{decrease}} (Fig~\ref{fig:conf-decrease}), \textbf{and \textit{recovery}} (Fig~\ref{fig:conf-recovery}). 

For each pattern, participants' distribution of self-efficacy scores varied; the \textit{stable} group had a mean self-efficacy of 5.95 ($SD=1.12$, $min=2$, $max=7$) with an average initial self-efficacy being 6.04 ($SD=1.07$, $min=2$, $max=7$), the \textit{decrease} group had a mean self-efficacy of 3.70 ($SD=1.67$, $min=1$, $max=7$) with an average initial self-efficacy being 6.00 ($SD=1.00$, $min=3$, $max=7$), and the \textit{recovery} group had a mean self-efficacy of 5.04 ($SD=1.52$, $min=1$, $max=7$) with an average initial self-efficacy being 6.12 ($SD=0.99$, $min=3$, $max=7$). 

To determine whether these trajectories were driven by baseline differences, we compared these initial scores using a Kruskal–Wallis test. The test showed no significant differences among the groups ($p>0.05$, not significant). This result shows that \textbf{participants’ initial self-efficacy levels did not differ across the three main trajectories}, suggesting that subsequent decreases or recoveries were not attributable to baseline differences.

For trust, among the same 302 participants, 76.5\% ($N=231$) were classified as \textit{stable}, 15.6\% ($N=47$) as \textit{increase}, 5.3\% ($N=16$) as \textit{recovery}, 1.7\% ($N=5$) as \textit{reversion}, and 1.0\% ($N=3$) as \textit{decrease}. Compared to self-efficacy, trust was predominantly stable, with net increases being more common than net decreases. Thus, \textbf{we focus on two major trajectory patterns for our subsequent analyses of trust:} \textbf{\textit{stable}} (Fig ~\ref{fig:trust-stable}) \textbf{and \textit{increase}} (Fig ~\ref{fig:trust-increase}).

Regarding score distributions, the \textit{stable} group had a mean trust of 6.25 ($SD=0.95$, $min=3$, $max=7$) and an average initial trust of 5.96 ($SD=0.93$, $min=3$, $max=7$), whereas the \textit{increase} group had a mean trust of 5.62 ($SD=1.45$, $min=1$, $max=7$) with a notably lower average initial trust of 4.00 ($SD=1.16$, $min=1$, $max=5$). 

To validate this disparity, we compared the initial trust scores between the two groups using a Mann–Whitney U test. The test revealed a significant difference ($U=9876$, $p<0.001$), showing that \textbf{trust increases occurred primarily among participants who began with lower initial trust}, whereas those who had higher trust remained stable.

\subsubsection{Interaction Between Self-Efficacy and Trust Over Time} \label{RQ1_2}
We present the results of mixed-effects models examining how self-efficacy and trust influence each other. The details of the regression models are explained in Section~\ref{method:mixed-effects}. We first report changes in self-efficacy over turns, followed by trust dynamics.

\paragraph{Changes in Self-Efficacy}
Through the mixed-effects model, we first found a significant main effect for \textit{turn} on self-efficacy ($\beta=-0.115$, $SE=0.023$, $z=-4.90$, $p<0.001$). On average, \textbf{participants' self-efficacy decreased over turns}, specifically, 0.115 points per turn on the 7-point Likert scale. We also found a significant main effect for \textit{trust} on self-efficacy ($\beta=0.100$, $SE=0.038$, $z=2.63$, $p<0.01$). This result suggests that \textbf{participants with a higher level of trust also maintained a higher overall self-efficacy} throughout the interactions.

We also found a significant \textit{turn} $\times$ \textit{trust} interaction ($\beta=0.038$, $SE=0.009$, $z=4.10$, $p<0.001$). This indicates that the effect of turns on self-efficacy was moderated by trust. In other words, \textbf{for participants with higher trust, the decrease in self-efficacy over turns was significantly smaller}, showing more resilient trajectories. This suggests that a high trust level may act as a buffer, mitigating the decrease in self-efficacy as the interaction progresses.

\paragraph{Changes in Trust}
In a complementary mixed-effects model, we found a significant main effect for \textit{turn} ($\beta=0.111$, $SE=0.011$, $z=10.13$, $p<0.001$) on trust. Unlike self-efficacy, the result suggests that \textbf{participants' trust increased over turns}, specifically, 0.111 points per turn on the 7-point Likert scale.

However, the effect of \textit{self-efficacy} on trust was not significant ($p>0.05$, not significant). Nevertheless, in the \textit{turn} $\times$ \textit{self-efficacy} interaction, we found a significance ($\beta=0.017$, $SE=0.004$, $z=3.67$, $p<0.001$). This shows that the effect of turns on trust was moderated by self-efficacy. Although self-efficacy was not a predictor of trust, \textbf{high self-efficacy showed a faster increase in trust over turns}, while lower self-efficacy showed a more gradual increase.

\subsection{RQ2. How are users’ prompting strategies associated with self-efficacy and trust trajectory patterns?} \label{RQ2}

\begin{figure*}[t!]
  \centering
  \newcommand{\colw}{0.49\linewidth}
  \makebox[\linewidth][c]{%
    \begin{subfigure}[t]{\colw}
      \includegraphics[width=\textwidth]{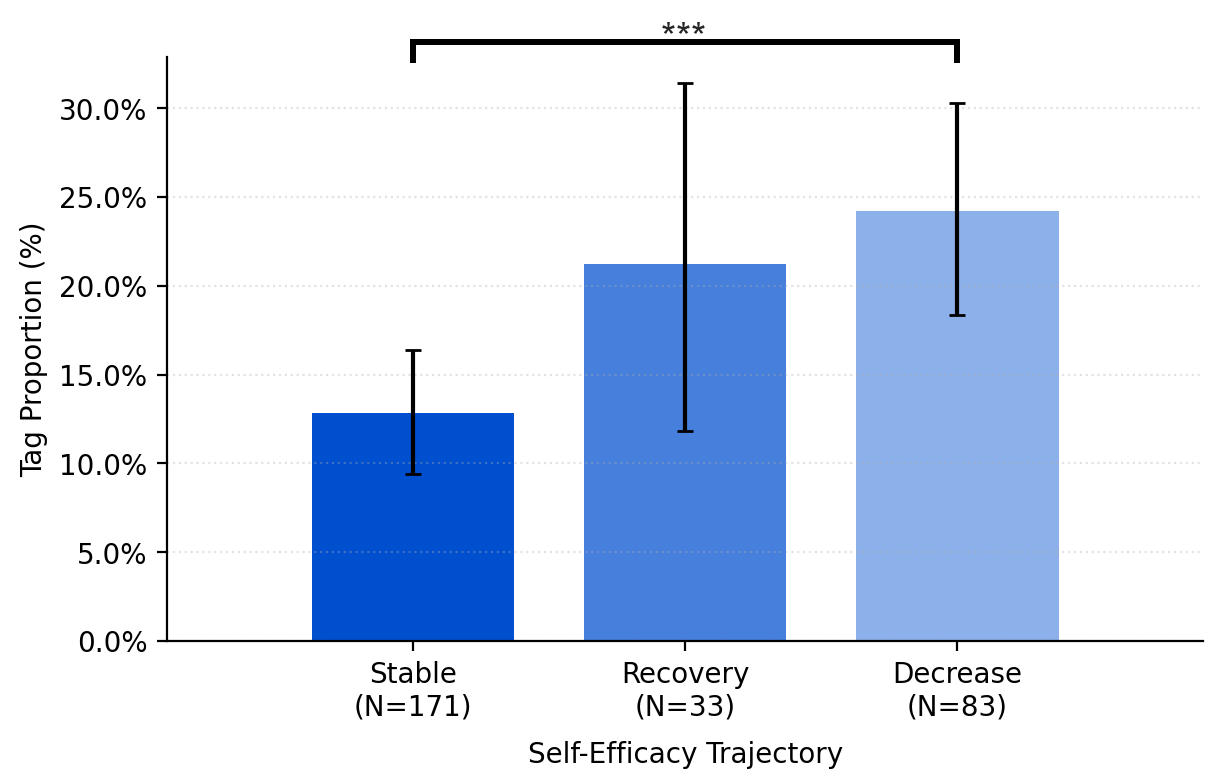}
      \caption{\textit{Editing} Prompt}\label{fig:editing}    
      \Description{Bar chart comparing the proportion of editing prompts across stable, recovery, and decrease self-efficacy trajectory patterns.}
    \end{subfigure}
    \hspace{0.04\linewidth}
    \begin{subfigure}[t]{\colw}
      \includegraphics[width=\textwidth]{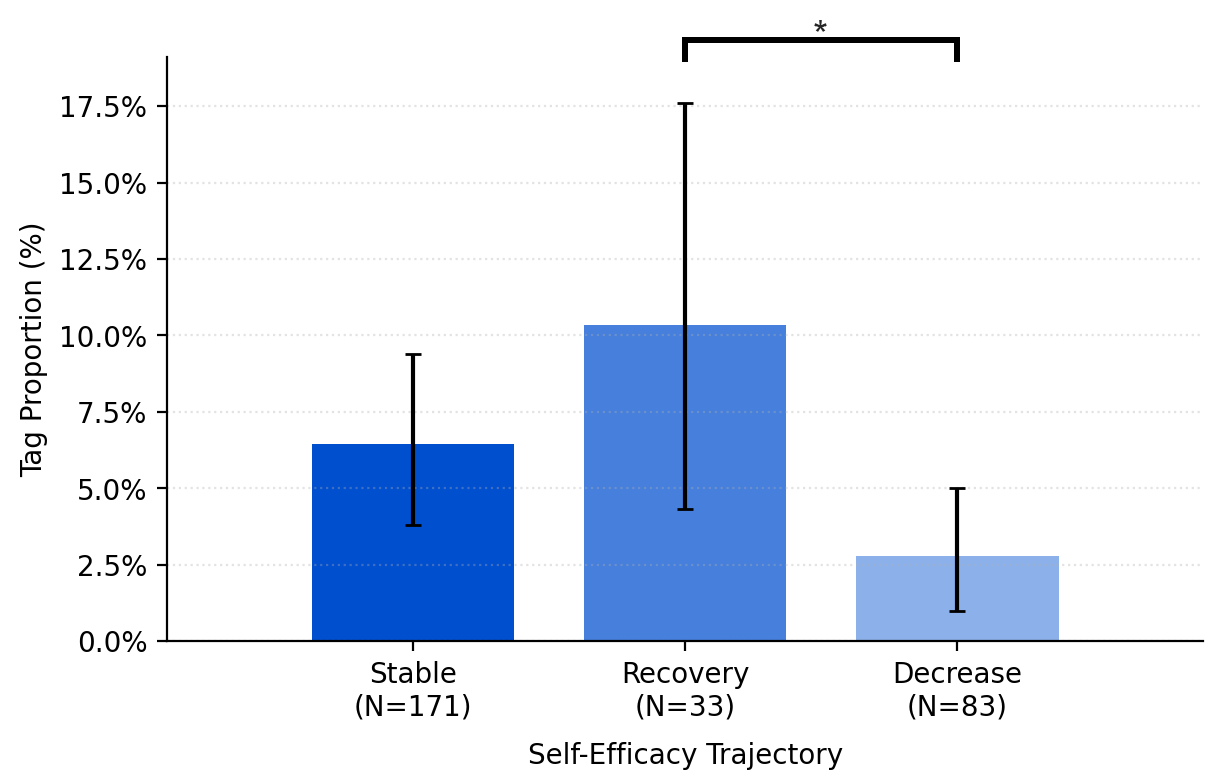}
      \caption{\textit{Reviewing} Prompt}\label{fig:review}    
      \Description{Bar chart comparing the proportion of reviewing prompts across stable, recovery, and decrease self-efficacy trajectory patterns.}
    \end{subfigure}%
  }
  \caption{Proportions of \textit{editing} and \textit{reviewing} prompts across different self-efficacy trajectory patterns. (*$p<0.05$, ***$p<0.001$)}
  \Description{Bar charts showing how the proportions of editing and reviewing prompts differ across three types of self-efficacy trajectory patterns.}
  \label{fig:prompt_individual}
\end{figure*}

We analyzed how self-efficacy and trust trajectory patterns relate to users' prompting strategies across two dimensions: (1) individual prompt usage and (2) prompt-to-prompt transitions, as explained in Section ~\ref{method:prompt} and ~\ref{method:strategy} respectively. To further contextualize these trajectory-level findings, we also conducted supplementary turn-level analyses of self-efficacy and trust across prompt categories. We present these additional findings in Appendix~\ref{appendix:RQ2_3}.

\subsubsection{Individual Prompt Usage} \label{RQ2_1}

\paragraph{Self-Efficacy}
We conducted Kruskal-Wallis tests to compare the proportional use of five main categories of prompts (\textit{drafting}, \textit{ideating}, \textit{editing}, \textit{information searching}, \textit{reviewing}) across three self-efficacy patterns (\textit{stable}, \textit{decrease}, \textit{recovery}). The result showed significant differences for \textit{editing} ($H(2)=13.57$, $p<0.001$) and \textit{reviewing} ($H(2)=6.71$, $p<0.05$). No significant differences were observed for the remaining categories ($p>0.05$, not significant).

We further conducted post-hoc comparisons for the significant prompt categories using Dunn's test with Holm-Bonferroni correction~\cite{holm1979simple} for multiple comparisons. We illustrate the pairwise comparison results of self-efficacy in Figure~\ref{fig:prompt_individual}.

We found that users with \textit{decrease} self-efficacy trajectory had a significantly higher proportion of \textit{editing} prompts than those in \textit{stable} trajectory ($p_{\text{adj}}<0.001$, Fig~\ref{fig:editing}). Moreover, users with \textit{recovery} trajectory showed a significantly higher proportion of \textit{reviewing} prompts than those in \textit{decrease} trajectory ($p_{\text{adj}}<0.05$, Fig~\ref{fig:review}). The other comparisons were not significant after correction. In other words, our results show that \textbf{users in \textit{decrease} trajectory request more for editing than those in \textit{stable}, while using fewer requests for reviewing than those in  \textit{recovery}}.

\begin{figure}
    \centering
    \includegraphics[width=\columnwidth]{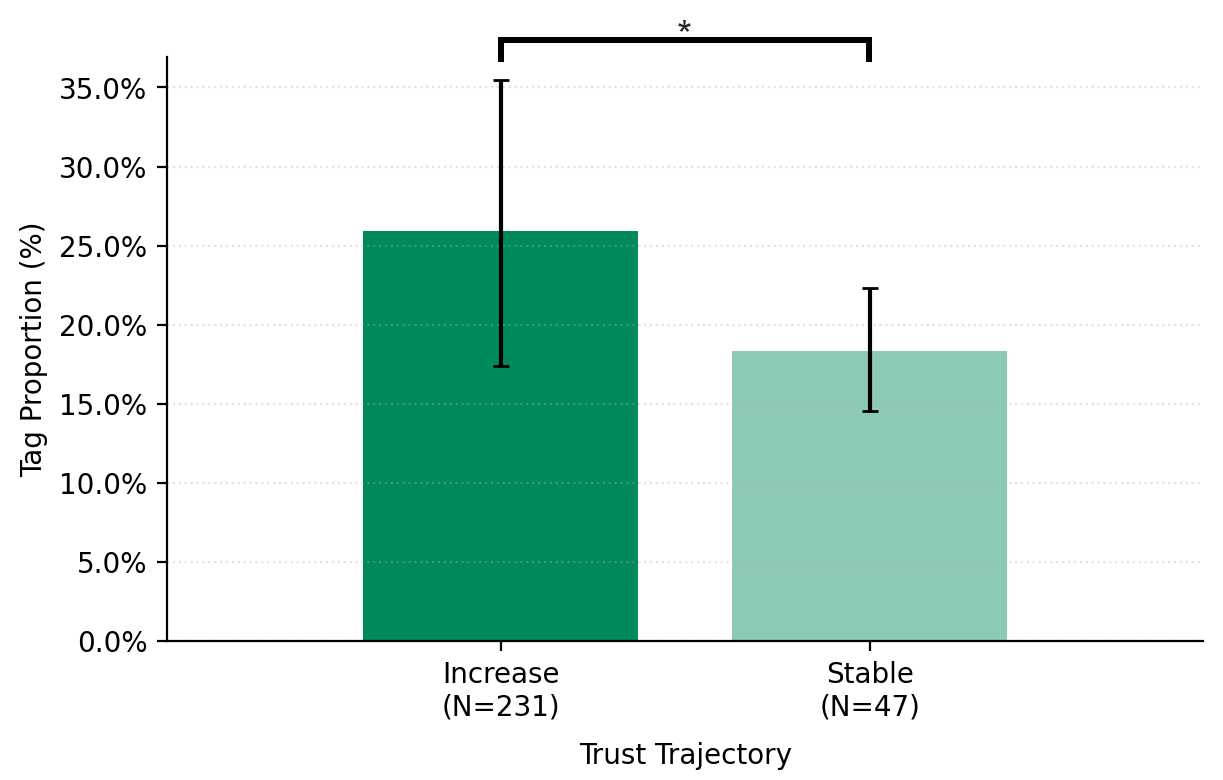}
    \caption{Proportions of \textit{information searching} prompts across different trust trajectory patterns. (*$p<0.05$)}\label{fig:information}
    \Description{Bar chart comparing the proportion of information searching prompts between participants with increasing trust trajectory patterns and those with stable trust trajectory patterns.}
\end{figure}

\paragraph{Trust}
We conducted Mann-Whitney U tests to compare prompt usage across two trust trajectory patterns (\textit{stable}, \textit{increase}). We found that \textbf{users in \textit{increase} trajectory request significantly more information searching than those in \textit{stable} trajectory} ($U=4293.5$, $p<0.05$, Fig~\ref{fig:information}). No significant differences were found for others. We illustrate the comparison results in Figure ~\ref{fig:information}.

\subsubsection{Prompt-to-Prompt Transitions} \label{RQ2_2}

\paragraph{Self-Efficacy}
Similarly, we conducted Kruskal-Wallis tests across self-efficacy patterns and found a significant difference for the \textit{drafting---editing} transition ($H(2)=16.55$, $p<0.001$) and \textit{editing---editing} transition ($H(2)=9.24$, $p<0.01$). 

\begin{figure*}[t!]
  \centering
  \newcommand{\colw}{0.49\linewidth}
  \makebox[\linewidth][c]{%
    \begin{subfigure}[t]{\colw}
      \includegraphics[width=\textwidth]{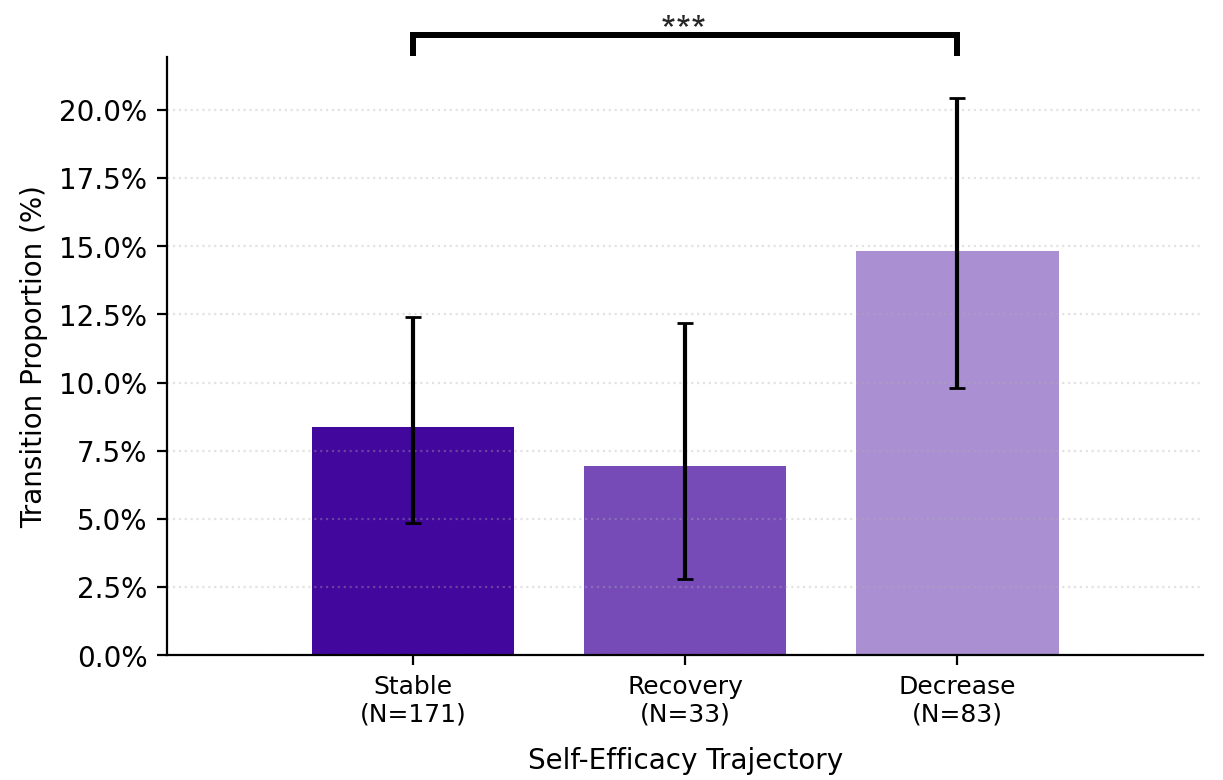}
      \caption{\textit{Drafting}---\textit{Editing} Transition}\label{fig:gen_edit}    
      \Description{Bar chart comparing the proportion of drafting-to-editing transitions across stable, recovery, and decrease self-efficacy trajectory patterns.}
    \end{subfigure}
    \hspace{0.04\linewidth}
    \begin{subfigure}[t]{\colw}
      \includegraphics[width=\textwidth]{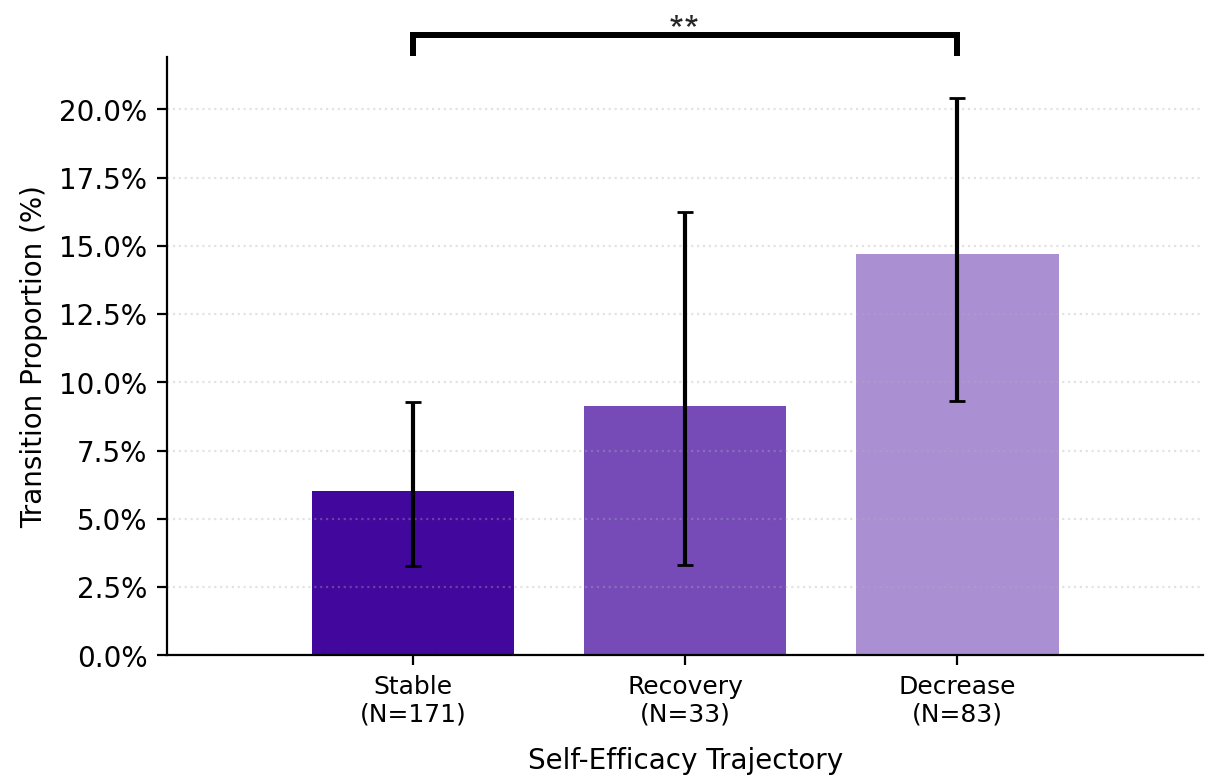}
      \caption{\textit{Editing}---\textit{Editing} Transition}\label{fig:edit_edit}    
      \Description{Bar chart comparing the proportion of editing-to-editing transitions across stable, recovery, and decrease self-efficacy trajectory patterns.}
    \end{subfigure}%
  }
  \caption{Proportion of \textit{drafting---editing} and \textit{editing---editing} transition across different self-efficacy trajectory patterns. (**$p<0.01$, ***$p<0.001$)}
  \Description{Bar charts showing how drafting-to-editing and editing-to-editing transitions vary across different self-efficacy trajectory patterns.}
  \label{fig:transition}
\end{figure*}

Post-hoc Dunn's test with Holm-Bonferroni correction~\cite{holm1979simple} further showed that \textbf{users with \textit{decrease} self-efficacy trajectory showed a significantly higher proportion of \textit{drafting---editing} transitions than those in the \textit{stable} trajectory} ($p_{\text{adj}}<0.001$, Fig~\ref{fig:gen_edit}). Similarly, \textbf{they also had a significantly higher proportion of \textit{editing---editing} transitions than the \textit{stable} trajectory} ($p_{\text{adj}}<0.01$, Fig ~\ref{fig:edit_edit}). Other pairwise comparisons were not significant. We illustrate these results in Figure ~\ref{fig:transition}. These findings reveal that users with decreasing self-efficacy exhibit a characteristic strategy of immediately transitioning from draft generation to editing, followed by consecutive editing behaviors.

\paragraph{Trust}
We conducted Mann-Whitney U tests for trust patterns across prompt transition types. However, we did not find any significant differences between them ($p>0.05$, not significant).

\subsection{RQ3. How are actual and perceived authorship associated with self-efficacy and trust trajectory patterns?} \label{RQ3}

We analyzed the relationship between the trajectory patterns and the authorship of the final outcome text. Note that actual authorship of each user was measured via lexical overlap and semantic similarity, and perceived authorship was measured via a post-survey on ownership and agency, as explained in Section~\ref{method:authorship}.

\paragraph{Self-Efficacy}
We conducted Kruskal-Wallis tests across three self-efficacy patterns (\textit{stable}, \textit{decrease}, \textit{recovery}) and found significant differences between them in terms of lexical overlap ($H(2)=12.65$, $p<0.01$), semantic similarity ($H(2)=12.77$, $p<0.01$), ownership ($H(2)=59.28$, $p<0.001$), and agency ($H(2)=25.93$, $p<0.001$).

\begin{figure*}[t]
  \centering
  \newcommand{\colw}{0.48\linewidth}
  \makebox[\linewidth][c]{%
    \begin{subfigure}[t]{\colw}
      \includegraphics[width=\textwidth]{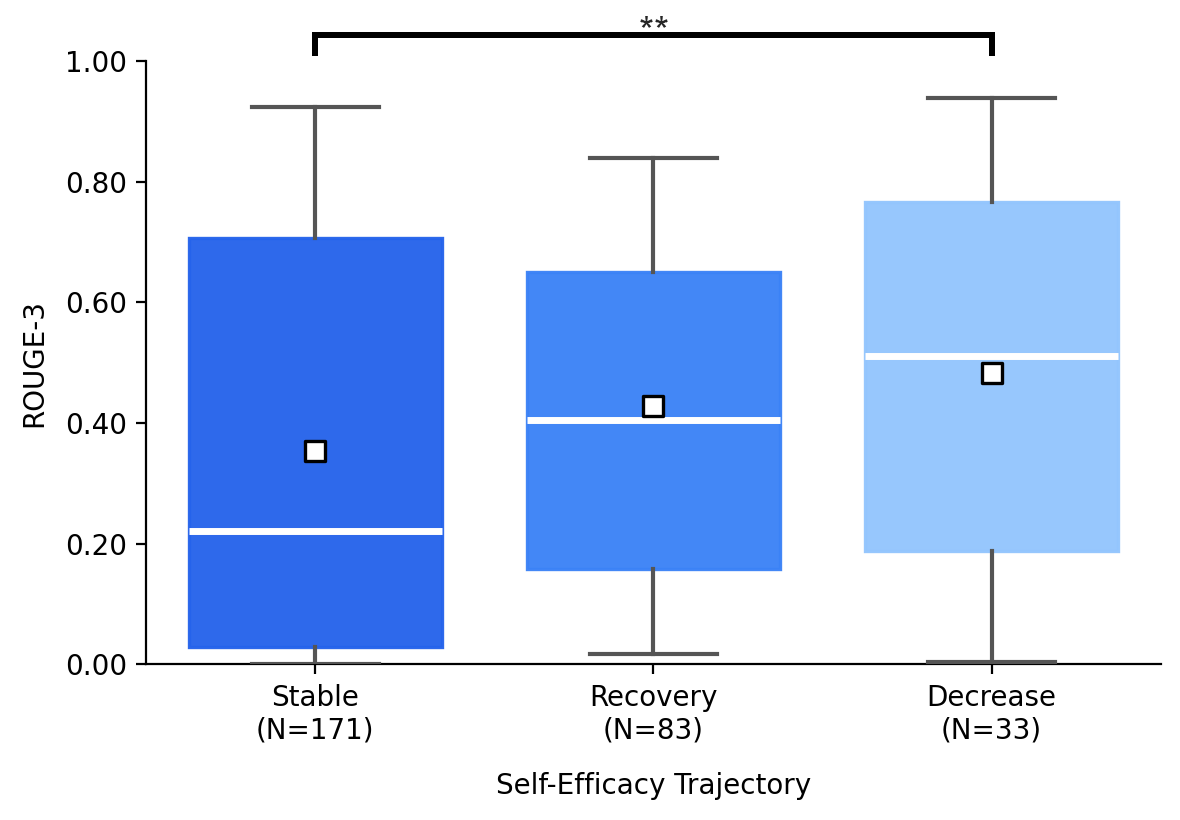}
      \caption{Lexical Overlap}\label{fig:rouge}    
      \Description{Boxplot comparing lexical overlap of participants’ writing across stable, recovery, and decrease self-efficacy patterns.}
    \end{subfigure}
    \hspace{0.04\linewidth}
    \begin{subfigure}[t]{\colw}
      \includegraphics[width=\textwidth]{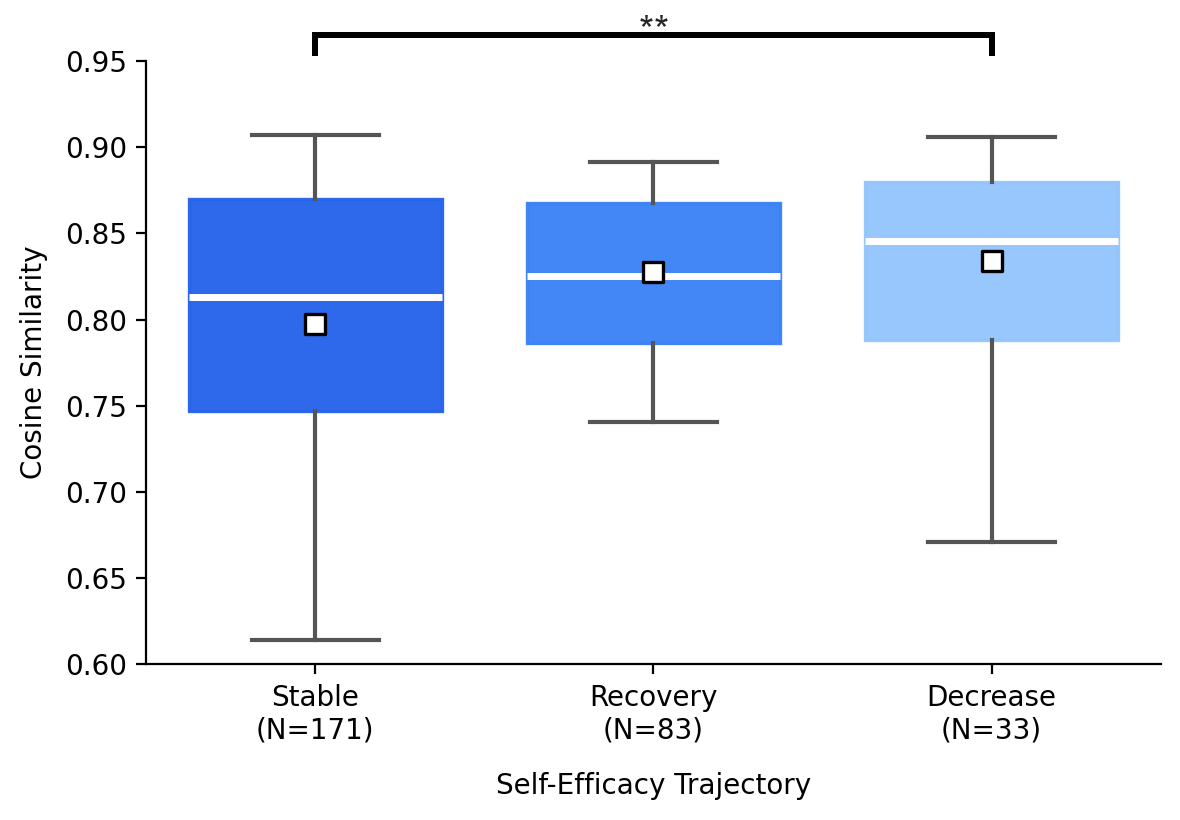}
      \caption{Semantic Similarity}\label{fig:embedding}    
      \Description{Boxplot comparing semantic similarity of participants’ writing across stable, recovery, and decrease self-efficacy patterns.}
    \end{subfigure}%
  }
    \par\vspace{1.5\baselineskip}
  \makebox[\linewidth][c]{%
    \begin{subfigure}[t]{\colw}
      \includegraphics[width=\textwidth]{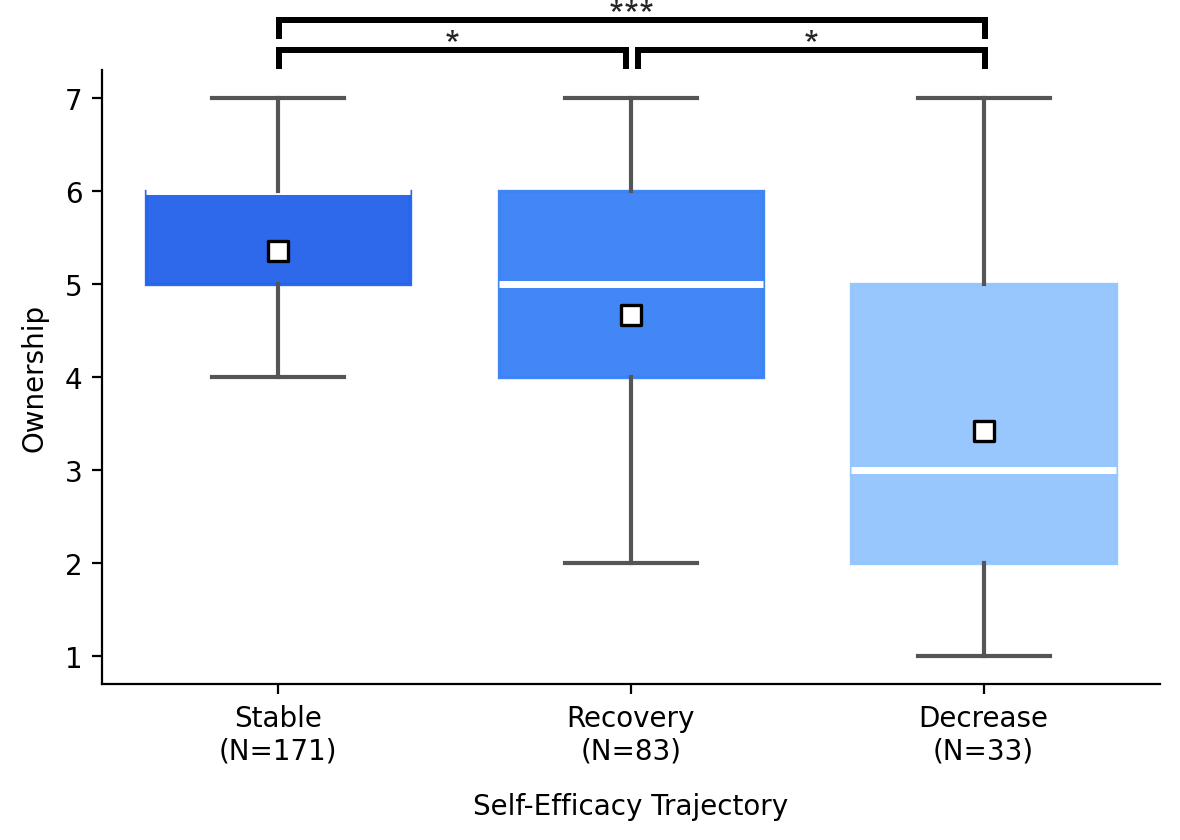}
      \caption{Ownership}\label{fig:ownership}    
      \Description{Boxplot comparing participants’ reported sense of ownership across stable, recovery, and decrease self-efficacy patterns.}
    \end{subfigure}
    \hspace{0.04\linewidth}
    \begin{subfigure}[t]{\colw}
      \includegraphics[width=\textwidth]{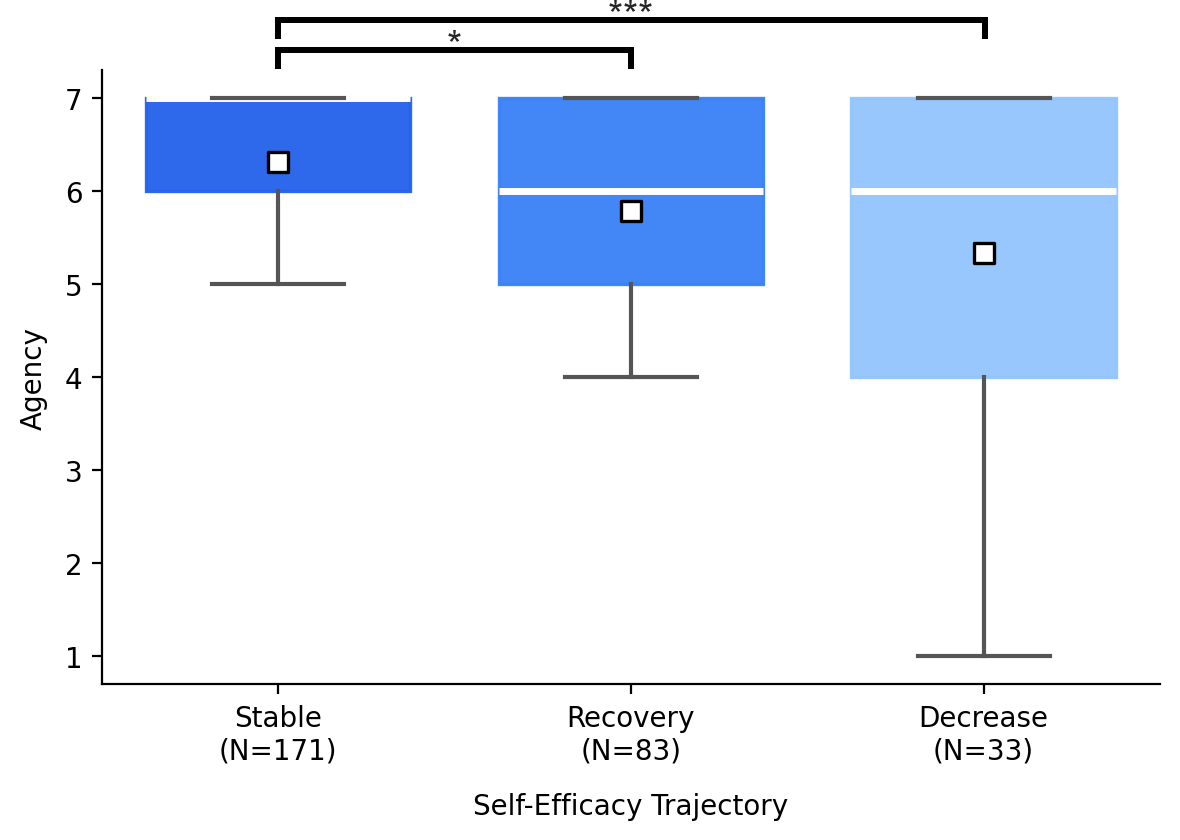}
      \caption{Agency}\label{fig:agency}    
      \Description{Boxplot comparing participants’ reported sense of agency across stable, recovery, and decrease self-efficacy patterns.}
    \end{subfigure}%
  }
  \caption{Actual (a–b) and perceived (c–d) authorship across different self-efficacy trajectory patterns. (*$p<0.05$, **$p<0.01$, ***$p<0.001$)}
  \Description{Four boxplots comparing authorship measures across stable, recovery, and decrease self-efficacy patterns. Subfigures (a) lexical overlap and (b) semantic similarity represent actual authorship, while subfigures (c) ownership and (d) agency represent perceived authorship.}
  \label{fig:authorship}
\end{figure*}

We further conducted post-hoc comparisons using Dunn's test with Holm-Bonferroni correction~\cite{holm1979simple} to correct for multiple comparisons. We illustrate the consequent pairwise results in Figure ~\ref{fig:authorship}.

For actual authorship, we found the \textit{decrease} trajectory had significantly higher lexical overlap than \textit{stable} ($p_{\text{adj}}<0.01$, Fig ~\ref{fig:rouge}). The other pairs were not significant after correction. Similarly, \textit{decrease} trajectory had significantly higher semantic similarity than \textit{stable} ($p_{\text{adj}}<0.01$, Fig ~\ref{fig:embedding}). The other pairs were not significant. 

For perceived authorship, we found the ownership significantly differed across all pairs (Fig ~\ref{fig:ownership}). Specifically, there was a clear hierarchy of \textit{stable} $>$ \textit{recovery} ($p_{\text{adj}}<0.05$), \textit{recovery} $>$ \textit{decrease} ($p_{\text{adj}}<0.05$), and \textit{stable} $>$ \textit{decrease} ($p_{\text{adj}}<0.001$). We found a similar pattern in agency (Fig ~\ref{fig:agency}); \textit{stable} trajectory had significantly higher agency than both \textit{recovery} ($p_{\text{adj}}<0.05$) and \textit{decrease} ($p_{\text{adj}}<0.001$). The \textit{recovery}---\textit{decrease} pair was not significant. 

Across the analyses, we observed a \textbf{consistent significant gap between self-efficacy trajectories of \textit{decrease} and \textit{stable}, where users in the \textit{decrease} trajectory show lower values for both actual and perceived authorship}. The \textit{recovery} trajectory was intermediate between the two, however, the differences were relatively smaller and often non-significant. 

\paragraph{Trust}
We conducted Mann-Whitney U tests across two trust patterns (\textit{stable}, \textit{increase}) and found no noticeable differences between them ($p > 0.05$, not significant) across every measure. 

%% file: sections/6_Discussion.tex
\section{Discussion}

In this section, we first interpret our study results. We then propose design implications for understanding and supporting authorship in human-LLM collaboration. Lastly, we discuss the possible generalizability of our findings to other domains.

\subsection{Interpretation of Results}
We discuss our interpretation of the results and their implications.

\subsubsection{Diverging Dynamics of Self-Efficacy and Trust}
We found that participants’ self-efficacy decreased while their trust increased over the course of the interaction (Section~\ref{RQ1_2}). The decline in self-efficacy may be attributed to upward social comparison with the model’s consistently high-quality outputs~\cite{bandura1977self}. As users repeatedly observed the system producing polished text more quickly than they could, this contrast may have highlighted gaps in their abilities. Prior work shows that performing alongside superior agents can erode self-efficacy through such comparisons, particularly when opportunities to demonstrate their own competence are limited~\cite{yaar2025s}. LLMs may amplify this dynamic by generating complete revisions instantly, leaving little room for users to formulate or refine text themselves. Even when users provide conceptual direction, the model executes the improvement, potentially reducing opportunities to develop skills or attribute outcomes to their efforts.

On the other hand, the increase in trust aligns with outcome-based trust calibration in automation~\cite{lee2004trust, wiczorek2019effects}, where trust evolves through the repeated observation of successful performance. In such calibration processes, users adjust trust by identifying reasoning flaws or system inaccuracies~\cite{tomsett2020rapid, yin2019understanding}. In writing, however, there is no binary correct answer, and evaluating revision quality requires significant cognitive effort. Hence, users rely on surface-level cues such as fluency and coherence~\cite{alter2009uniting, reber2010epistemic}, with each fluent output perceived as a success. This leads users to overweight the model's apparent competence, driving trust accumulation. These dynamics create an asymmetric structure: users receive frequent confirmation of the system's competence but little confirmation of their own. This imbalance may explain why trust and self-efficacy diverged during collaboration, suggesting that the capabilities making LLMs powerful writing partners may simultaneously shift users' perceptions of their own competence and contribution.

\subsubsection{Role of Prompting Strategies in Self-Efficacy Trajectories}
Our findings show that users whose self-efficacy declined relied heavily on \textit{editing} requests while those who recovered self-efficacy used more \textit{review} requests (Section~\ref{RQ2}). This pattern can be understood through findings from the educational domain, showing that direct corrective feedback positions learners as passive recipients, while diagnostic feedback that identifies issues maintains learner agency and supports self-efficacy~\cite{griffiths2023can}. These studies emphasize that effective feedback should provide users the ability to maintain control over revision decisions in determining when, where, and how to implement suggested changes.

Heavy reliance on editing requests reflects a dynamic where users position LLMs as the authoritative agent responsible for text improvements, effectively delegating revision decisions to the model. This can reduce their own agency as opportunities to make and justify revision decisions themselves become limited. On the other hand, a shift toward review-oriented interactions indicates users who maintain their position as primary authors while leveraging the LLMs as a diagnostic tool for feedback. This approach preserves writer agency, consistent with research showing that peer review can reduce anxiety while building self-efficacy~\cite{busby2023writing}. Therefore, understanding how users perceive the LLM's role as a collaborator is important, and our work shows how these perceptions can be understood through their prompting behaviors.

\subsection{Design Implications}

We present design implications for understanding and supporting authorship in human-LLM collaboration.

\subsubsection{Conceptualizing Self-Efficacy as a Dynamic Process}

Our findings suggest that self-efficacy operates as a dynamic process during human-LLM collaboration rather than a static trait. Prior research on self-efficacy and authorship has shown that higher self-efficacy aligns with stronger authorship and authorial identity~\cite{pittam2009student, maguire2013self}, whereas lower self-efficacy is associated with greater susceptibility to plagiarism~\cite{fatima2020impact}. However, existing work largely treats self-efficacy as a static, between-person attribute that is typically measured at a single point in time.

In contrast, our findings reveal that within-session trajectories of self-efficacy provide a more informative basis for understanding authorship loss. While participants' initial self-efficacy scores did not differ across trajectory groups (Section~\ref{RQ1_1}), within-session declines and fluctuations in self-efficacy were significantly associated with decreased authorship (Section~\ref{RQ3}). Specifically, users who experienced declining self-efficacy not only lost actual authorship by adopting more of the LLM's direct phrasing and ideas but also reported weaker perceived ownership and agency over their final text. Moreover, users whose self-efficacy recovered showed intermediate authorship outcomes between the stable and decreasing groups, a pattern that a simple pre–post comparison would obscure.

Taken together, these findings suggest that understanding authorship in human–LLM collaboration requires accounting for how self-efficacy evolves over the course of writing. Conceptualizing self-efficacy as a dynamic trajectory offers a more nuanced lens for examining when and why authorship is strengthened or weakened, and provides a basis for assessments that can, in turn, inform the design of interventions to preserve user authorship.

\subsubsection{Using Behavioral Signals to Examine Self-Efficacy Shifts}
Self-efficacy decline can undermine both actual and perceived authorship (Section~\ref{RQ3}), making it important to understand when and how these shifts occur. Our findings suggest that user behavior during collaboration can serve as unobtrusive signals of changes in self-efficacy, revealing moments in which authorship may become vulnerable. Specifically, we identified two key behavioral patterns: prompting strategies and how users incorporate LLM outputs.

Prompting strategies strongly correspond to self-efficacy trajectories (Section~\ref{RQ2}). A high frequency of consecutive editing prompts or rapid shifts from drafting to editing may indicate a decline in self-efficacy, whereas an increase in reviewing prompts may reflect efforts to regain a sense of control. Such shifts can be detected using methods like prompt intent classification~\cite{bodonhelyi2024user}, providing a way to identify moments of potential authorship loss.

Moreover, how users integrate LLM outputs provides complementary insight into self-efficacy shifts. As self-efficacy declines, users tend to adopt a larger portion of the model’s suggestions (Section~\ref{RQ3}). Tracking how semantic similarity or lexical overlap between user-written text and model-generated text changes over time can reveal shifts from active authorship toward passive acceptance. Techniques such as keystroke logging~\cite{leijten2013keystroke} enable capturing these incorporation dynamics, offering additional evidence of when users’ contributions begin to recede.

Together, these signals offer a foundation for examining self-efficacy trajectories in situ and open up opportunities to explore strategies that help sustain authorship when self-efficacy is at risk.

\subsubsection{Opportunities for Supporting Authorship in LLM-Assisted Writing} \label{discussion:support}
Our findings on behavioral signals and self-efficacy trajectories open up several directions for designing systems that better support user authorship in LLM-assisted writing. We discuss opportunities for designing environments that can foster a more intentional and reflective collaboration, helping users maintain active agency rather than drifting into passive acceptance.

One opportunity lies in helping users stay aware of their interaction patterns and shift toward more reflective engagements over time. Our results revealed that users with declining self-efficacy tended to fall into repetitive ``editing loops'' of requesting direct revisions (Section~\ref{RQ2_2}), while those who regained self-efficacy more often sought evaluation rather than relying on the model to fix their text (Section~\ref{RQ2_1}). Since prior work suggests that making people's own thoughts visible can foster metacognitive awareness~\cite{tankelevitch2024metacognitive}, systems could help users recognize their engagement patterns by making collaboration history more visible~\cite{hoque2024hallmark} or steering responses toward review and reflection~\cite{yuan2024generative}. Such interventions may help users move from passive acceptance of suggestions toward more intentional collaboration, with greater agency over their writing.

Another opportunity lies in supporting users' information needs during the writing process. Our results show that information-seeking behaviors were associated with increasing trust in the LLM (Section~\ref{RQ2_1}), which in turn buffered against declines in self-efficacy (Section~\ref{RQ1_2}). Notably, trust increases occurred primarily among participants who began with lower initial trust (Section~\ref{RQ1_1}), suggesting that informational support may be especially beneficial during early stages of interaction. Building on these findings, systems could provide relevant background information or supporting evidence when users need it, an approach shown to improve writing quality~\cite{koskela2018proactive} while helping users remain in control of the writing process. However, as LLMs may occasionally retrieve irrelevant content or hallucinate sources~\cite{press2024citeme, li2024attributionbench}, systems could consider strategies to maintain calibrated trust, such as signaling uncertainty~\cite{kocielnik2019will} or prompting users to validate sources~\cite{kim2025fostering}.

\subsection{Generalization of Results}
Our study examined how users' self-efficacy and trust evolve during LLM-assisted writing. While trust in the LLM increased over time, users' self-efficacy decreased. We discuss how this pattern can generalize to other cognitively demanding, open-ended tasks where LLMs can be particularly valuable in providing diverse assistance.

Prior work on AI-assisted tools, particularly in creative domains, has largely focused on how LLMs can enhance efficiency~\cite{yuan2022wordcraft}, performance~\cite{urban2024chatgpt}, and user satisfaction~\cite{wang2025aideation}. However, our findings point to a different dimension of AI-assisted work, indicating that users’ self-efficacy in their abilities and sense of authorship may decline during collaboration with LLMs. This highlights the need for future research to examine not only what AI tools can produce, but also how they affect users' psychological states over time. These considerations may become especially salient in multimodal AI systems, where distinguishing individual contributions is more challenging than in text-based collaboration.

Our findings are also particularly significant in educational contexts, where the consequences of declining self-efficacy and losing authorship can be extremely severe. In learning environments, students' belief in their own abilities is fundamental to academic motivation and long-term development. Recent work has already shown that AI overreliance can diminish students' critical thinking and engagement~\cite{zhai2024effects}. Our observed decline in self-efficacy may represent a crucial pathway to these detrimental outcomes, limiting students' capacity to approach challenging tasks independently.

Both creative and educational contexts are examples of a broader category of cognitively demanding tasks where tension between increased performance and diminished self-efficacy is likely to emerge. Thus, our implications can be extended beyond these domains to collaborative AI systems supporting cognitively intensive work.

%% file: sections/7_Limitation.tex
\section{Limitations and Future Work} \label{limitation}

We acknowledge several limitations of our study and suggest potential future work.

First, our measures of self-efficacy and trust relied on self-reported ratings collected after each interaction turn with the LLM. Although most participants indicated that this process did not affect their interactions, a subset reported that it was distracting and made them more self-conscious about their choices (Section~\ref{method:study}). Despite our efforts to design the ratings to be as lightweight as possible, this suggests that it may still have influenced participants' natural interaction patterns in subtle ways, even among those who reported no disruption. Specifically, participants who experienced the ratings as distracting may have partially shifted their work to external tools (e.g., other LLMs), but because our logging was limited to in-platform events (Section~\ref{method:interface}), we cannot determine whether this occurred. To better capture users’ natural interactions, future work could explore less intrusive approaches, such as inferring self-efficacy and trust from behavioral signals---including typing latency and editing pauses~\cite{zhang2021using}, dwell time~\cite{shi2022effects}, or eye gaze~\cite{wu2025releyeance}. Alternatively, expanding data collection to include pageview and cross-platform activity could enable post-hoc filtering of sessions involving external tool use, improving data integrity.

Second, our findings identified empirical patterns in how self-efficacy and trust evolved within an interaction session with the LLM, but not the qualitative reasons behind them. To uncover why these shifts occur, future work could employ qualitative methods like user interviews or think-aloud protocols to capture the moment-to-moment rationale behind users' decisions. Bridging the gap between observed behaviors and subjective goals would provide actionable guidance for designing effective interventions to preserve user authorship (Section~\ref{discussion:support}). Furthermore, although we analyzed patterns within a multi-turn interaction, the study was confined to a single session. This limits our understanding of whether the observed shifts are temporary or develop into long-term changes. To capture how these dynamics change, future research could extend our design to a longitudinal study that tracks user perceptions across multiple sessions, revealing how repeated interactions shape users' evolving relationship with AI assistance.

Finally, this study did not account for individual or system-related factors that prior work has shown to influence self-efficacy and trust, such as task proficiency~\cite{qian2024take} or variations in system tone and expression~\cite{zhou2024rel, kim2024m}. Our analysis also used a single LLM, so we could not examine how such stylistic and behavioral differences across models might alter user perceptions~\cite{fanous2025syceval}. Future research could extend this work by studying more diverse user groups and varying LLM configurations to better understand how these contextual factors interact with the dynamics of self-efficacy and trust.

%% file: sections/8_Conclusion.tex
\section{Conclusion}
We investigated how self-efficacy and trust evolve during LLM-assisted writing through an empirical study with 302 participants. We found that while trust generally increased throughout interaction, self-efficacy tended to decline, buffered by high trust. Trajectory-level analyses revealed that users with decreasing self-efficacy were more likely to rely on passive editing strategies and showed diminished authorship, whereas those with recovering self-efficacy engaged more in review-oriented interactions with intermediate authorship outcomes between the decreasing and stable groups. Our findings show that self-efficacy operates as a dynamic process that critically shapes how people collaborate with LLMs, highlighting the need to examine and respond to within-session shifts to preserve user authorship. We call for future work to move beyond single-session self-reports, exploring longitudinal studies and unobtrusive measures to better understand long-term trajectories of self-efficacy, trust, and human agency in AI-assisted work.

%% file: sections/10_Appendix.tex
\clearpage
\appendix
\section{Procedure for Essay Grading}\label{appendix:grading}

We generated initial essay scores using a prompt adapted from the automated essay scoring prompts of Yoo et al.~\cite{yoo2025dress} with the official College Board rubric for the AP English Language and Composition Argument Essay\hyperref[fn:cb]{\textsuperscript{\ref*{fn:cb}}}. The rubric evaluates three criteria for a total of six points: (1) thesis (0-1), (2) evidence and commentary (0-4), and (3) sophistication (0-1). We instructed \texttt{gpt-5-nano}\footnote{\url{https://platform.openai.com/docs/models/gpt-5-nano}} to generate scores with brief justifications for each criterion, enabling the author to review and finalize the grades. The detailed prompt is shown below.

% \vspace{0.3cm}
\begin{framed}
\noindent
You are grading an AP English Language and Composition Argument Essay. \\ \\
\#\#\# Rubric: \\
\texttt{<rubric\_criteria>} \\ \\
Comments: For each row, provide a justification of the score in this exact bullet format: \\
- Thesis: [your justification here] \\
- Evidence \& Commentary: [your justification here] \\
- Sophistication: [your justification here]
\\ \\
\#\#\# Output Instructions: \\
Return ONLY a valid JSON object with the following keys: \\ 
\{thesis: integer (0-1), evidence\_commentary: integer (0-4), sophistication: integer (0-1), comment: string\} 
\\ \\
\#\#\# Essay Prompt: \\
\texttt{<prompt\_text>} \\ \\
\#\#\# Student Essay: \\
\texttt{<essay\_text>} 
\end{framed}
% \vspace{0.3cm}

After obtaining the initial rubric-based scores, one of the authors iteratively reviewed the essays and adjusted these scores to determine final bonus recipients. Since many essays received similar high scores, each iteration involved manually evaluating top-ranked essays, adjusting scores to better differentiate quality, and re-ranking all entries. This process was repeated until the top-performing subset was clearly distinguished and stabilized. In total, the author adjusted 6.10\% of the LLM-generated scores, and the top 34 participants (11.26\%) received the bonus payment.

\section{Turn-Level Analysis of Self-Efficacy and Trust by Prompt Category}\label{appendix:RQ2_3}

This appendix reports supplementary turn-level analyses examining how users' self-efficacy and trust ratings differed across prompt categories during the interaction. These analyses provide additional context for interpreting the trajectory-level findings of Section~\ref{RQ2} by revealing how each prompt category is related to users' momentary ratings at the turn level.

\subsection{Self-Efficacy}
\begin{figure}[h]
  \includegraphics[width=\columnwidth]{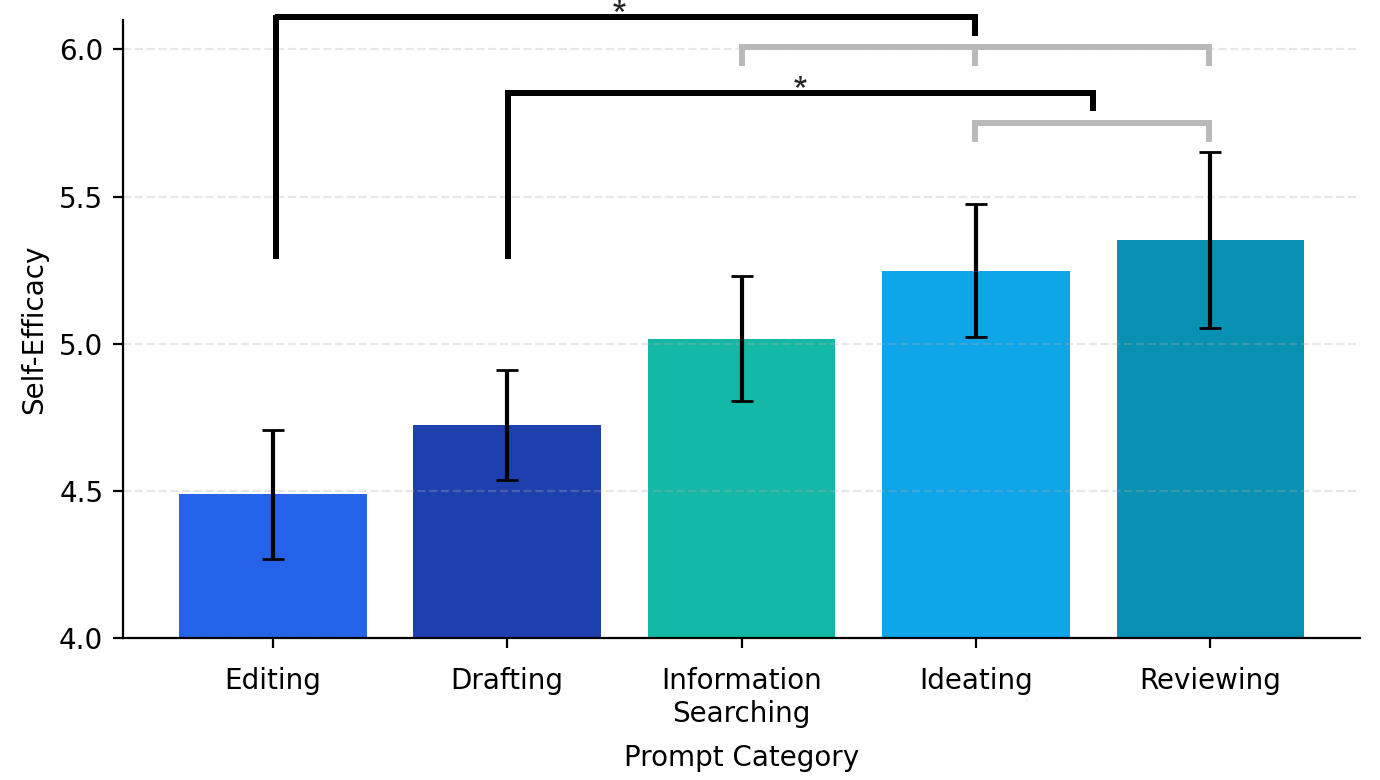}
  \caption{Turn-level self-efficacy ratings across prompt categories (*$p<0.05$)}\label{fig:efficacy_tags}    
  \Description{Bar chart showing mean turn-level self-efficacy ratings for each prompt category (editing, drafting, information searching, ideating, and reviewing).}
\end{figure}

We conducted a Kruskal–Wallis test to compare turn-level self-efficacy across the five main prompt categories and found significant differences among them ($H(4)=28.74$, $p<0.001$). Post-hoc Dunn tests with Holm-Bonferroni correction~\cite{holm1979simple} revealed that \textit{editing} prompts were associated with significantly lower self-efficacy than \textit{ideating}, \textit{information searching}, and \textit{reviewing} prompts ($p_{\text{adj}}<0.05$ for all). Moreover, \textit{drafting} prompts also showed lower self-efficacy than \textit{ideating} and \textit{reviewing} ($p_{\text{adj}}<0.05$ for all). We illustrate this result in Figure~\ref{fig:efficacy_tags}.

These turn-level findings align with our trajectory-level findings in Sections~\ref{RQ2_1} and~\ref{RQ2_2}. Users in the \textit{decrease} trajectory relied heavily on \textit{editing} prompts and engaged in \textit{drafting---editing} or \textit{editing---editing} transitions, while those in the \textit{recovery} trajectory made more \textit{reviewing} requests. The turn-level results suggest that \textit{editing} and \textit{drafting} are more often used when users feel less efficacious, whereas \textit{ideating}, \textit{information searching}, and \textit{reviewing} tend to occur when self-efficacy is relatively higher. This indicates that heavy reliance on editing both reflects and potentially reinforces lower-efficacious trajectories, while reviewing appears characteristic of users who maintain or regain their sense of capability.

\subsection{Trust}
\begin{figure}[h]
  \includegraphics[width=\columnwidth]{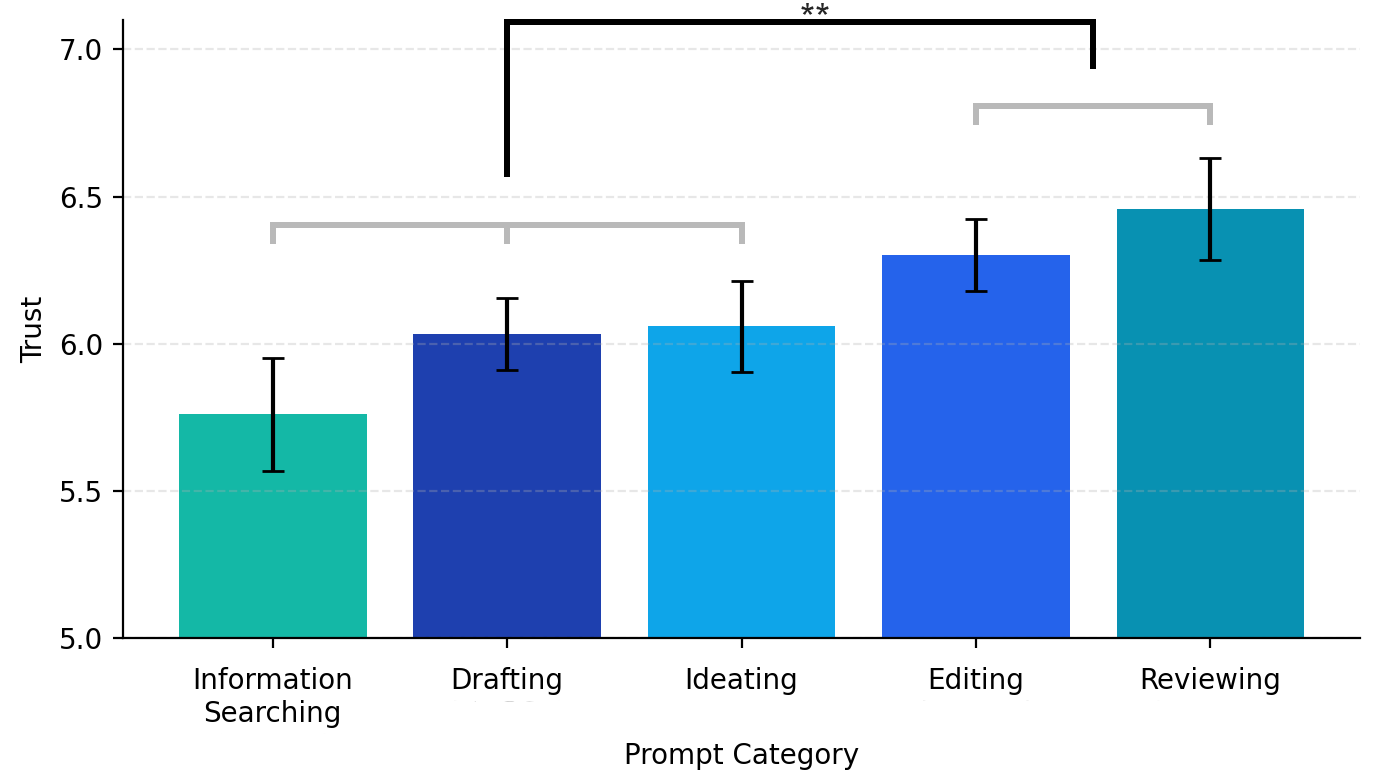}
  \caption{Turn-level trust ratings across prompt categories (**$p<0.01$)}\label{fig:trust_tags}    
  \Description{Bar chart showing mean turn-level trust ratings for each prompt category (information searching, drafting, ideating, editing, and reviewing).}
\end{figure}

We also conducted a Kruskal–Wallis test for trust ratings across prompt categories and found significant differences ($H(4)=39.02$, $p<0.001$). Post-hoc Dunn tests with Holm-Bonferroni correction~\cite{holm1979simple} revealed that \textit{reviewing} and \textit{editing} prompts were associated with higher trust than the other three ($p_{\text{adj}}<0.01$ for all), while \textit{information searching} showed the lowest mean trust (\textit{M=5.76}). We illustrate this result in Figure~\ref{fig:trust_tags}.

These offer contextual insight into the trajectory-level findings of Section~\ref{RQ1_1} and ~\ref{RQ2_1}. Users in the \textit{increase} trajectory began with lower initial trust and more frequently used \textit{information searching} prompts, and the turn-level analysis similarly shows that \textit{information searching} turns coincide with lower trust. In contrast, \textit{editing} and \textit{reviewing} prompts, which involve working more directly with the outcome text, are associated with higher trust. Taken together, these patterns suggest that users tend to deal with lower trust through information-seeking interactions, and that once trust is higher, they are more likely to request edits and reviews that directly shape the outcome text.

\subsection{Considerations for Interpretation}
It is important to acknowledge that this turn-level analysis aggregates data across all interactions without normalizing for the varying number of turns per participant. Consequently, participants who engaged in more turns contribute proportionally more data points to each prompt category, potentially amplifying the influence of highly active users. As a result, these findings may partly reflect individual interaction styles or strategies rather than generalizable properties of the prompt categories themselves. While this analysis offers supportive evidence that complements our trajectory-level findings, we emphasize that these associations serve as an exploratory context for interpreting the trajectory patterns rather than definitive evidence regarding intrinsic properties of each prompt category.